\documentclass[12pt]{article}

\usepackage{amsmath}
\usepackage{amssymb}
\usepackage{epsfig}
\usepackage{graphicx}
\usepackage{cite}

\newcommand{\capdef}{}
\newcommand{\mycaption}[2][\capdef]{\renewcommand{\capdef}{#2}%
        \caption[#1]{{\footnotesize #2}}}
\makeatletter
\renewcommand{\fnum@table}{\textbf{\tablename~\thetable}}
\renewcommand{\fnum@figure}{\textbf{\figurename~\thefigure}}
\makeatother

\newcounter{myenumi}

\renewcommand{\themyenumi}{\roman{myenumi}}
{\end{list}}

\newlength{\myem}
\newcounter{mysubequation}[equation]

\makeatletter
\renewcommand{\section}{\@startsection{section}{1}{0em}{-\baselineskip}%
{\baselineskip}{\normalfont\large\bfseries}}
\renewcommand{\subsection}%
{\@startsection{subsection}{2}{0em}{-0.7\baselineskip}%
{0.7\baselineskip}{\normalfont\bfseries}}
\makeatother

\newcommand{\bi}{\begin{itemize}}
\newcommand{\ei}{\end{itemize}}

\newcommand{\be}{\begin{equation}}
\newcommand{\ee}{\end{equation}}
\newcommand{\bea}{\begin{eqnarray}}
\newcommand{\eea}{\end{eqnarray}}

\newcommand{\ldm}{\Delta m_{31}^2}

\newcommand{\deltacp}{\delta_{\mathrm{CP}}}
\newcommand{\stheta}{\sin^22 \theta_{13}}

\newcommand{\ie}{{\it i.e.}}

\newcommand{\eg}{{\it e.g.}}

\newcommand{\cf}{{\it cf.}}

\newcommand{\etc}{{\it etc.}}
\newcommand{\eq}{Eq.}

\newcommand{\fig}{Fig.}

\newcommand{\Ref}{Ref.}
\newcommand{\Refs}{Refs.}
\newcommand{\Sec}{Sec.}

\newcommand{\Tab}{Table}

\newcommand{\equ}[1]{\eq~(\ref{equ:#1})}
\newcommand{\figu}[1]{\fig~\ref{fig:#1}}

\begin{document}

\begin{titlepage}

\renewcommand{\thefootnote}{\alph{footnote}}

\vspace*{-3.cm}
\begin{flushright}
\end{flushright}


\renewcommand{\thefootnote}{\fnsymbol{footnote}}
\setcounter{footnote}{-1}

{\begin{center}
{\large\bf
Which long-baseline neutrino experiments are preferable?
} \end{center}}
\renewcommand{\thefootnote}{\alph{footnote}}

\vspace*{.3cm}
{\begin{center} {\large{\sc
                V.~Barger\footnote[1]{\makebox[1.cm]{Email:}
                barger@physics.wisc.edu},
                P.~Huber\footnote[2]{\makebox[1.cm]{Email:}
                phuber@physics.wisc.edu},
                D.~Marfatia\footnote[3]{\makebox[1.cm]{Email:}
                marfatia@ku.edu}, and
                W.~Winter\footnote[4]{\makebox[1.cm]{Email:}
                winter@physik.uni-wuerzburg.de}
                }}
\end{center}}
\vspace*{0cm}
{\it
\begin{center}

\footnotemark[1]$^,$
\footnotemark[2]
       Department of Physics, University of Wisconsin,
       Madison, WI 53706, USA

\footnotemark[3]
       Department of Physics \& Astronomy, University of Kansas,
       Lawrence, KS 66045, USA

\footnotemark[4]
       Institut f{\"u}r theoretische Physik und Astrophysik, Universit{\"a}t W{\"u}rzburg, \\
       D-97074 W{\"u}rzburg, Germany

\end{center}}

\vspace*{1.5cm}

{\Large \bf
\begin{center} Abstract \end{center}  }

We discuss the physics of superbeam upgrades, where we focus on T2KK,
a NuMI beam line based experiment NO$\nu$A*, and a wide band beam (WBB) experiment
independent of the NuMI beam line. For T2KK, we find that the
Japan-Korea baseline helps resolve parameter degeneracies, but the
improvement due to correlated systematics between the two detectors (using
identical detectors) is only moderate.  
For an upgrade of NO$\nu$A with a liquid argon detector,
we demonstrate that the Ash River site is preferred compared to
alternatives, such as at the second oscillation maximum, and is the
optimal site within the U.S.  For a WBB experiment, we find
that high proton energies and long decay tunnels are preferable. 
We compare water Cherenkov and liquid argon technologies, and find the break-even point
in detector cost at about 4:1. In order to compare the physics potential 
of the different experimental configurations,
we use the concept of {\it exposure} to normalize the performance.
We find that experiments with WBBs are the best experimental
concept. NO$\nu$A* could be competitive with sufficient luminosity.
If $\stheta > 0.01$, a WBB experiment can perform better than a neutrino factory. 
 

\vspace*{.5cm}

\end{titlepage}

\newpage

\renewcommand{\thefootnote}{\arabic{footnote}}
\setcounter{footnote}{0}

\section{Introduction}

The motivation for measuring all the parameters of the neutrino mass matrix
is that such knowledge will allow us to sift through the many neutrino
mass models available~\cite{Mohapatra:2006gs} 
and lead us to a better understanding of the origin of
neutrino masses.

To make further progress in our exploration of the neutrino sector
precision experiments are needed.
The MINOS experiment~\cite{Ables:1995wq} will provide an accurate
determination of $|\ldm|$ in the near future~\cite{Barger:2001yx,Huber:2004ug}.
Reactor experiments with multiple detectors, like the Double CHOOZ~\cite{Ardellier:2006mn} 
and Daya Bay~\cite{guo:2007ug} experiments will either constrain the angle
$\theta_{13}$ (which mixes the solar and atmospheric oscillation scales)
from above, or from below if $\stheta$ is larger than about 0.03~\cite{Huber:2006vr}.
 Also, the Tokai-to-Kamioka (T2K)
experiment~\cite{Itow:2001ee} which is under construction, and expected to be in operation in a
couple of years, will detect $\nu_\mu \to \nu_e$ oscillations if
$\theta_{13}$ is favorably large, but due to its relatively
short baseline will not be
sensitive to the neutrino mass hierarchy.
While the reactor and T2K experiments
will provide important guidance about $\theta_{13}$, they are inherently
limited in their abilities to determine the mass hierarchy and to tell us
if the CP symmetry is broken in the lepton sector. For these measurements
upgraded experiments with neutrino beams are necessary. See Ref.~\cite{Barger:2003qi}
for a review.

In this paper, we discuss upgraded experiments with high intensity
neutrino beams resulting from proton beams
with target powers above 1 MW that are directed towards
either very large or very sophisticated
detectors. We compare the potential of these experiments 
with that of a neutrino factory and a $\beta$-beam experiment. 
Our discussion, in the usual parlance, is that of a ``Phase II''
program, and is addressed to experts. 

The goals of our study are
\begin{enumerate}
\item To determine for the Tokai-to-Kamioka-and-Korea (T2KK) 
experiment~\cite{Ishitsuka:2005qi,Hagiwara:2005pe,Kajita:2006bt} (consisting of
two identical detectors of half
the size as the originally envisaged  megaton-scale detector), 
\subitem (a) the effect of placing half the fiducial mass at the 1050~km baseline
compared to having all the fiducial mass at 295~km. See Section 3.2.
\subitem (b) how it helps that the two detectors are identical,
\ie, that the systematics are fully correlated. See Section 3.3.
\item To determine the optimal location for a second detector
in an upgraded NuMI off-axis experiment, which we dub NO$\nu$A*. See Sections 4.2-4.3.
\item To determine the optimal baseline, 
proton energy and decay tunnel length for a Fermilab-based wide band beam. See Sections 5.2-5.3.
\item To compare the setups of different proposed superbeam experiments on an
equal-footing by expressing their sensitivities as functions of exposure, as in 
Ref.~\cite{Barger:2006kp}, and to assess if it is better to place a detector on-axis or off-axis.
See Section 6.1.
\item To quantify the robustness of our conclusions under variations of 
exposure, systematics and the value of $|\ldm|$. See Section 6.2.
\item To determine the smallest value of $\theta_{13}$ for which superbeam experiments are 
competitive with a neutrino factory or a $\beta$-beam experiment. See Section 6.3.

\end{enumerate}

\section{Analysis Techniques}
\label{sec:pheno}

There is an eight-fold degeneracy~\cite{Barger:2001yr} in the determination of 
the oscillation parameters that arises from three two-fold 
degeneracies~\cite{Barger:2001yr,Burguet-Castell:2001ez,Minakata:2001qm,Fogli:1996pv}:
\begin{itemize}
\item Intrinsic ($\theta_{13}\,,\deltacp$)
  degeneracy with ($\theta_{13}\,,\deltacp$) $\rightarrow$
  ($\theta_{13}'\,,\deltacp'$).
\item Sign-degeneracy with $\Delta m^2_{31}\rightarrow -\Delta m^2_{31}$.
\item Octant-degeneracy with $\theta_{23}\rightarrow\pi/2-\theta_{23}$.
\end{itemize}
The octant-degeneracy will not influence our results and discussions
since we set the true value for the atmospheric mixing angle to the
current best-fit value $\sin^22\theta_{23}=1$.
The intrinsic degeneracy is very
often not explicitly present (as a disconnected degenerate solution) for superbeams,
but it appears as a strong correlation (connected degenerate solutions). In most cases,
the sign-degeneracy affects the performance. For example,
 the CP-conserving solutions of the
wrong hierarchy solution may destroy the CP violation (CPV) sensitivity because
CP conservation cannot be excluded.
Longer baselines and therefore strong matter
effects~\cite{Barger:1980tf} 
may help to resolve this degeneracy and to improve the physics performance, because
they may intrinsically separate the different hierarchy solutions; see, \eg,
\Refs~\cite{Barger:2001yr,Minakata:2001qm,Minakata:2002qi,Minakata:2002qe,Minakata:2003ca,Winter:2003ye}
for a pictorial representation in terms of bi-probability diagrams. Therefore, a detector
located at a sufficiently long baseline is usually a key element of any superbeam upgrade
to break the remaining degeneracy.

In order to reduce correlations, especially between $\theta_{13}$ and $\deltacp$,
different strategies
are possible for superbeam upgrades. For experiments using the off-axis technology, a
second detector at a different location can provide complementary information for a
different $L$ and/or $E$. In addition, better energy
resolution and higher statistics may provide measurements of the transition probability
 for different values
of $E$. Alternatively, a wide energy spectrum can provide
these measurements even with a rather poor energy resolution.
In this study, we discuss several very different such approaches and demonstrate
how these strategies affect the physics performance.

For the quantitative analysis, we use the
GLoBES software~\cite{Huber:2004ka,Huber:2007ji}. As input, or so-called true
values, we use, unless stated otherwise (see \Refs~\cite{Maltoni:2004ei,Schwetz:2006dh})
\begin{eqnarray}
\Delta m^2_{31} & =& 2.5\cdot10^{-3}\,  \mathrm{eV}^2 \, , \quad\sin^2\theta_{23}=0.5 \, , \nonumber \\
\Delta m^2_{21} & = &8.0 \cdot10^{-5}\, \mathrm{eV}^2 \, , \quad\sin^2\theta_{12}=0.3 \, ,
\label{equ:params}
\end{eqnarray} where $\stheta \lesssim 0.1$.
In anticipation of near future experiments, 
we assume a 4\% external measurement of solar oscillation parameters  (see, \eg,
\Ref~\cite{Minakata:2004jt,Bandyopadhyay:2004cp}), a 10\% measurement of the
atmospheric parameters (see, \eg, \Ref~\cite{Antusch:2004yx,Huber:2004ug}), and include matter density uncertainties
of the order of 5\%~\cite{Geller:2001ix,Ohlsson:2003ip} uncorrelated
between different baselines.

In order to make an unbiased comparison of the physics potentials of the
experimental setups
we will in some cases 
consider their sensitivities as functions of {\it {exposure}}~\cite{Barger:2006kp}
\begin{equation}
\label{equ:exposure}
\mathcal{L}=  \mathrm{detector \, mass
\, [Mt] \times target \, power \, [MW] \times running \, time \, [10^7 \, s]} \, .
\end{equation}
The exposure is a measure of the integrated luminosity.
For the Fermilab-based experiments, we use $1.7 \cdot 10^7$
seconds uptime per year, and for T2KK, we use $10^7$ seconds uptime per year 
(as per the corresponding documents). 

\section{T2KK: An Off-Axis Experiment with Two Identical Detectors}

T2KK is a Japanese-Korean approach for a superbeam upgrade of the T2K 
 off-axis superbeam experiment~\cite{Itow:2001ee}.
Originally, it was planned to upgrade T2K to T2HK (Tokai-to-HyperKamiokande) 
with a $4 \, \mathrm{MW}$ proton beam and a megaton-size water Cherenkov detector. 
Recently, it was recognized that placing a part of the detector mass in Korea would 
enhance the mass hierarchy and degeneracy resolution 
potential~\cite{Ishitsuka:2005qi,Hagiwara:2005pe,Kajita:2006bt}. In particular, it 
was emphasized in \Ref~\cite{Ishitsuka:2005qi} that using identical detectors and the 
same off-axis angle would reduce the impact of systematics significantly. More recent 
studies have investigated the off-axis angle optimization and the splitting of the 
detector mass between the different sites~\cite{Hagiwara:2006vn,Hagiwara:2006nn}. However, in 
this case, the detectors cannot be assumed to be identical anymore in the sense that 
systematics are correlated between the two detectors. In this study, we focus on the
 setup with correlated detectors because this aspect is very specific to the T2KK 
idea and has certain characteristics. 

\subsection{Simulation Details}

Details of our simulation of T2KK including the beam spectrum, cross sections, 
and efficiencies is based on the setup ``JHF-HK'' from \Ref~\cite{Huber:2002mx} 
with an off-axis angle of $2^\circ$. In order to include the systematics 
correlation between the two detectors, we use the GLoBES 
software~\cite{Huber:2004ka,Huber:2007ji} which allows for user-defined systematics.
 For detector masses and running times, we follow \Ref~\cite{Ishitsuka:2005qi}. 
We use four years of neutrino running, followed by four years of antineutrino running. 
We assume $52 \cdot 10^{20} \, \mathrm{pot/yr}$ in either mode corresponding to a target 
power of $4 \, \mathrm{MW}$. The (identical) water Cherenkov detectors of fiducial mass 
$270 \, \mathrm{kt}$ will be located 
at a distance of $295 \, \mathrm{km}$ and $1050 \, \mathrm{km}$ with the same off-axis angle. 
According to our definition of the exposure (\cf, \equ{exposure}), this corresponds to 
$\mathcal{L} = 17.28 \, \mathrm{Mt \, MW \, 10^7 \, s}$. 
We follow \Ref~\cite{Ishitsuka:2005qi} for the systematics modeling as well,
\ie, we assume that the systematics are completely correlated between the two identical 
detectors in Japan and Korea, and the uncorrelated errors are 
small.
We include the signal and background normalization errors of the appearance channels 
(typically 5\% each), and allow for an independent 5\% background energy 
calibration error. In addition, we include the disappearance channels with a 
2.5\% normalization uncertainty and a 20\% background normalization 
uncertainty~\cite{Huber:2002mx}. Note that we assume the errors to be completely 
correlated between the two detectors, but to be completely uncorrelated 
between the neutrino and
antineutrino channels (as opposed to \Ref~\cite{Ishitsuka:2005qi}). 
Our simulation uses the spectral information for the quasi-elastic (QE) charged-current (CC)
events only (because the detector cannot measure the hadronic energy deposition for the 
non-QE events), and the total rate for all CC events.

\begin{figure}[t]
\begin{center}
\includegraphics[width=\textwidth]{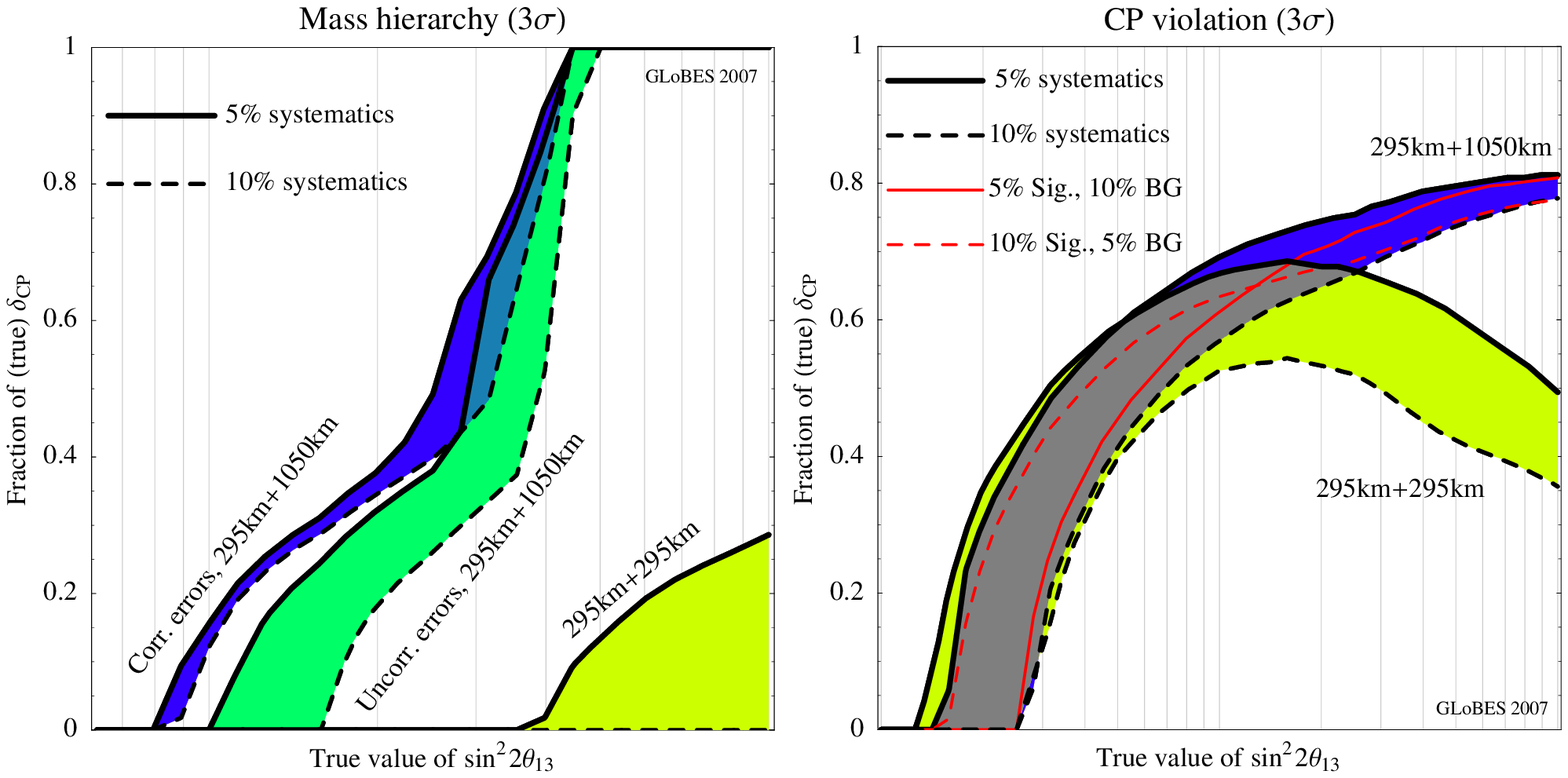}
\end{center}
\mycaption{\label{fig:t2kk} Impact of systematics, Japan-Korea baseline, and detector correlation: 
 $3 \sigma$ sensitivity to the (normal) mass hierarchy and CP violation
 for T2KK as a function of $\stheta$ and the fraction of $\deltacp$; for a definition see
Fig.~3 of Ref.~\cite{Barger:2006vy}.
In the figures, several different systematics options and
baseline combinations are shown, where ``systematics'' refers to both signal and
background normalization errors of the appearance channel. 
For the dark gray-shaded (blue) and light gray-shaded (yellow) regions, systematics are assumed to be fully correlated between the detectors, which corresponds to only
one detector for the $295 \, \mathrm{km}+295 \, \mathrm{km}$ option. For the medium gray-shaded region (left plot), systematics are assumed to be uncorrelated between two detectors. We
do not show this region in the right plot because it is almost identical to the dark-shaded
(correlated) region. In both plots, regions of overlap have grayscale shading intermediate to 
those of the overlapping regions. Note that we assumed  
neutrino and antineutrino systematics to be uncorrelated.
}
\end{figure}

\subsection{Role of Systematics Correlation and Second Baseline}

We present our main results in \figu{t2kk}. This figure show the mass hierarchy
and CP violation measurements for several systematics and baseline options, where
the total detector mass is $0.27 \, \mathrm{Mt} + 0.27 \, \mathrm{Mt}$.
The ``systematics'' correspond to the signal and background normalization errors
of the appearance channels, which we have determined to be the main impact factors.

A determination of the mass hierarchy will be difficult without the
$1050 \, \mathrm{km}$ baseline. As one can deduce from
\figu{t2kk} (left), a large systematic error of 10\% would even destroy the measurement
completely for the short baseline only. However, for the combination of baselines,
the impact of systematics is small (dark/blue region). We also show the curves for
uncorrelated errors between the two detectors (medium gray/green region). Having
identical detectors clearly helps for the mass hierarchy measurement. 
Not only is the absolute $\stheta$ reach better for correlated systematics, but also 
the impact of  worse systematics becomes
smaller. We have tested that a 2\% systematic uncertainty would not substantially 
improve the results. 
It is the background normalization error which affects the
results most because it dominates the measurement for small $\stheta$.

For CP violation, the situation is very different (\cf, \figu{t2kk}, right).\footnote{The 
fraction of (true values of) $\deltacp$
for which an experiment is sensitive to CPV cannot be unity even
for the best setup and very large $\stheta$. The reason is that one cannot establish CP
violation at and around the CP conserving values which one wants to discriminate from
the CP violating effect. Therefore, the fraction of $\deltacp$ has to be smaller than one by
definition.
This indicator describes how close to CP conservation one can establish CP
violation, \ie, how small CP violating effects an experiment can find.} The long baseline
helps to resolve the degeneracies for large $\stheta$, while for
small $\stheta$ having all the detector mass at $295 \, \mathrm{km}$ would be better because 
of the larger statistics. 
The impact of the absolute systematics is strong and is independent of the baseline combination. 
We do not show the curves
for uncorrelated errors in the right panel of \figu{t2kk} because they overlap significantly 
with the correlated curves. This means that while the longer baseline is clearly beneficial, 
the fact that the detectors are identical is not. 
It will be very important to control
the systematics for CP violation measurements since the shorter
baseline practically measures CP violation, and the systematics impact is
largest between the uncorrelated neutrino and antineutrino errors. 
Note that the impact of different individual systematic uncertainties is illustrated by the 
gray curves in the right panel of \figu{t2kk}. 
For small $\stheta$, the background normalization dominates, while for 
large $\stheta$ the signal normalization has the greatest impact. 

\subsection{Impact of Degree of Correlation}

\begin{figure}[t]
\begin{center}
\includegraphics[width=\textwidth]{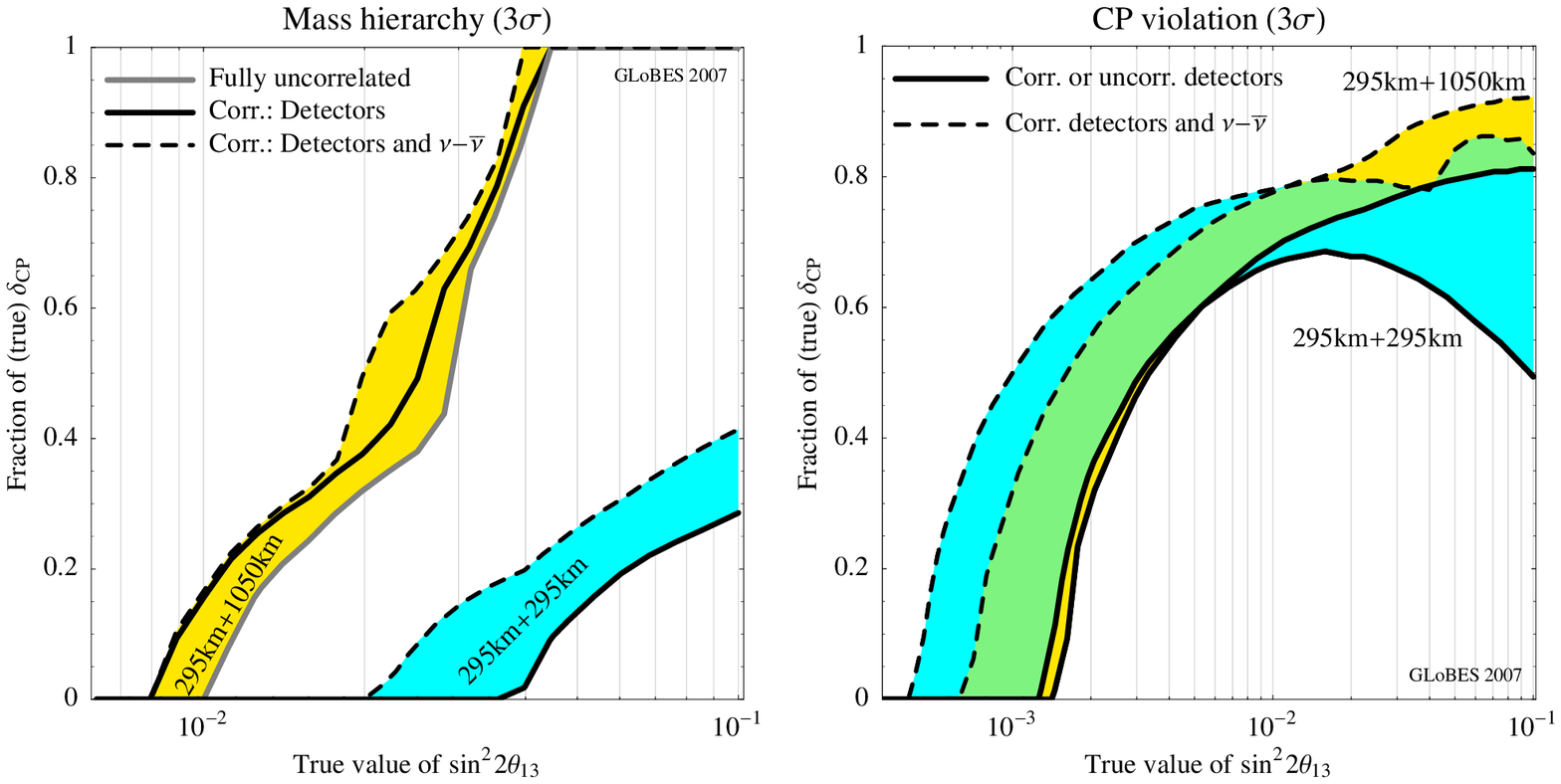}
\end{center}
\mycaption{\label{fig:t2kkcorr} Impact of detector versus neutrino-antineutrino correlation: 
$3 \sigma$ sensitivity to the (normal) mass hierarchy and CP violation
 for T2KK as a function of $\stheta$ and the CP fraction. 
In the figures, several different systematics options and
baseline combinations are shown. The systematics can be either fully uncorrelated between the two detectors and between neutrinos and antineutrinos (gray curves), fully correlated between the
two detectors, but uncorrelated between neutrinos and antineutrinos (black curves), or fully correlated between the two detectors and neutrinos and antineutrinos (dashed curves), where 5\% signal and background normalization uncertainties are chosen for all curves. The bands reflect the possible outcome for 5\% systematics depending on the achievable degree of correlation. Note that in the right plot, the curves for fully uncorrelated detectors are practically identical to the black curves (not shown) and the region where the bands overlap has grayscale 
shading intermediate to 
those of the overlapping bands.
For the 295~km+295~km option, the correlated detectors correspond to one 
detector with uncorrelated systematics.}
\end{figure}

In order to compare our results with those of \Ref~\cite{Ishitsuka:2005qi}, 
we illustrate the impact of different degrees of correlation in \figu{t2kkcorr}. 
Systematics can be either fully uncorrelated between the two detectors 
and between neutrinos and antineutrinos (gray curves), fully correlated between the two 
detectors, but uncorrelated between neutrinos and antineutrinos (black curves), or fully 
correlated between the two detectors and neutrinos and antineutrinos (dashed curves). 
We have chosen 5\% signal and background normalization uncertainties. 
If there is some correlation between neutrino and antineutrino systematics, 
especially the CP violation measurement will be improved. 
In fact, we have tested that this correlation makes the CP violation sensitivity 
rather insensitive to the absolute magnitude of the systematic errors, whereas the 
correlation between the detectors hardly has any impact. The bands in \figu{t2kkcorr} reflect 
all possible outcomes for 5\% systematics depending on the achievable degree of correlation. 
For the rest of this study, we adopt the conservative point of view that systematics between 
the neutrino and antineutrino modes are uncorrelated (but correlated between the two detectors). 
Because they are operated at different times with different beams, the cross section errors 
are certainly uncorrelated, and the background reduction may also be different (\eg, 
at least the $\pi^0$ production rates are different). In addition, the argument of some 
correlation between neutrino and antineutrino systematics could be applied to other experiments 
as well, and we treat all experiments with comparable assumptions.
Eventually, note that while correlated systematics between neutrinos and antineutrinos 
would increase the physics potential, it would relatively weaken the physics case for the 
Japan-Korea baseline because all detector mass at $L=295 \, \mathrm{km}$ would be better 
for CP violation measurements at small $\stheta$ and improve the mass hierarchy potential, too.

In summary, we find that the $1050 \, \mathrm{km}$ baseline clearly helps for both
mass hierarchy and CP violation measurements regardless of the size of the systematic errors.
However, it is only useful that the detectors are identical (and have completely correlated 
systematics) for the mass hierarchy measurement -- as long as neutrino and antineutrino 
systematics are assumed to be uncorrelated. In addition, we have tested that matter density
uncertainties and the assumptions for the solar parameter precisions practically do not have an 
impact. For the rest of this study, we use the standard T2KK setup with 5\% systematics 
correlated between the two detectors, but uncorrelated between the neutrino and antineutrino 
running; \cf, \Tab~\ref{tab:setups}.

\section{NO$\boldsymbol{\nu}$A*: A New Experiment Based on the NuMI Beam Line}
\label{sec:nova}

The first experiment based on the NuMI beam line at Fermilab is the MINOS 
experiment~\cite{Michael:2006rx}. It has been proposed to use the NuMI beam line for a 
future off-axis superbeam experiment. The NuMI off-axis experiment (NO$\nu$A) 
was originally designed with a Low-Z-tracking calorimeter to be placed off-axis in the 
NuMI beam~\cite{Ayres:2002nm}. Since the then preferred baseline of $L
\simeq 712 \, \mathrm{km}$ turned out to be too short to be
complementary to the T2K experiment in Japan, longer baselines were
suggested in
\Refs~\cite{Barger:2002rr,Barger:2002xk,Huber:2002rs,Minakata:2003ca}.\footnote{In
  principle, one can also adjust the off-axis angle
  instead~\cite{Mena:2006uw} in order to adjust the energy.} As a rule
of thumb,
\begin{equation}
\left( \frac{L}{E} \right)_{\mathrm{NuMI}} \gtrsim \left( \frac{L}{E} \right)_{\mathrm{T2K}} \label{equ:le}
\end{equation}
was found in order to yield synergistic physics, which translates to
$L_{\mathrm{NuMI}} \gtrsim 862 \, \mathrm{km}$ for $0.72^\circ$
off-axis angle. Hence, a longer baseline, $L \simeq 810
\, \mathrm{km}$ to the Ash River site in Minnesota, has been proposed for the NO$\nu$A
experiment~\cite{Ayres:2004js}. A typical off-axis angle suggested is 0.85$^\circ$,
corresponding to 12~km off-axis at this baseline. In addition, a
Totally Active Scintillating Detector (TASD) is the now accepted
detector technology, often considered with a mass of $25 \,
\mathrm{kt}$. Alternative sites with longer baselines in Canada
(because of the beam geometry), such as Vermillion Bay, with
potentially attractive physics potential are not actively being considered.

For upgrades of the NO$\nu$A experiment, several
ideas have been proposed. Two approaches are
discussed in the NO$\nu$A proposal~\cite{Ayres:2004js}.
The luminosity could be increased by using a proton
driver~\cite{Albrow:2005kw}, or a second detector could be placed at the second
oscillation maximum at a short baseline $L = 710 \, \mathrm{km}$ 
with a larger off-axis angle, 2.4$^\circ$. 
We refer to this setup as ``2nd maximum''. In principle,
increasing the protons on target has the same effect as increasing the detector
mass at the same site. However, a different detector technology may
allow for better energy resolution, background rejection, \etc, which
needs to be explored. Another possible upgrade is to use the
same $L/E$ for both detectors, an idea similar to that encapsulated in \equ{le}. Since
longer baselines may not be possible with a detector within the U.S., a
larger off-axis angle of 2.4$^\circ$ combined with a short baseline $L
\simeq 200 \, \mathrm{km}$, called ``Super-NO$\nu$A'', was suggested
in \Refs~\cite{MenaRequejo:2005hn,Mena:2005ri}. In this section, we
investigate the issue of site optimization using a liquid argon
detector as a second detector. Note that for the NuMI-based experiment,
no new beam line is needed,
but the existing beam line constrains the combinations of
allowed baselines and off-axis angles. 

\subsection{Simulation Details}
\label{sec:novadetails}

Our simulation of the NO$\nu$A experiment is based on that in
\Ref~\cite{Huber:2002rs} updated with the numbers from
\Ref~\cite{Ayres:2004js}. We assume $10^{21}$ protons on target per
year, corresponding to a thermal target power of $1.13 \, \mathrm{MW}$
(for both the neutrino and antineutrino mode), which may be achieved
in the Super-NuMI phase at Fermilab~\cite{fnalprotons}. For phase~I,
we assume a 25~kt TASD located at a baseline of $810 \, \mathrm{km}$
and 0.85$^\circ$ (12~km) off-axis.\footnote{The very specific detector
  mass of phase~I does not affect our results, because phase~I
  is only used for the optimization of the phase~II detector location.
  We do not expect a major impact of a smaller phase~I detector on the
  optimization discussion.} The fluxes for phase~I are taken
from~\cite{MessierF}. This location corresponds to the Ash River site.
For the running times, we assume three years of neutrino running and
three years of antineutrino running, followed by another period of
three years of neutrino running and three years of antineutrino
running, \ie, 12 years of operation altogether. For phase~II, which we
refer to as NO$\nu$A*, we assume a 100~kt liquid argon time
projection chamber (LArTPC) operated with the same proton luminosity
for three years of neutrino running, followed by three years of
antineutrino running. Since the optimization of the site for the
second detector is our primary goal in this section, we do not choose
a specific detector location.  For the LArTPC
simulation~\cite{Fleming}, we assume an overall signal efficiency of
$0.8$. The energy response of a LArTPC depends on the event type. Since the
type of event can be unambiguously determined, we
split the event sample into QE events, which have an energy
resolution of $5\%/\sqrt E$, and all other CC events, which have
a $20\%/\sqrt E$ energy resolution. We furthermore assume that all neutral-current (NC) 
events will be identified as such, which means that only the
beam intrinsic $\nu_e$'s and $\bar\nu_e$'s remain as backgrounds. We
adopt the NuMI beam fluxes for the low-energy option from
\Ref~\cite{MessierF}, and we use the interpolation routines provided
together with the fluxes to obtain the flux at any site. The
interpolation errors are smaller than 10\%~\cite{MessierF} and should
not impact our results. The systematic errors are assumed to be
5\% for the signal and background normalizations.

\subsection{Optimal Detector Location for NO$\boldsymbol{\nu}$A*}
\label{sec:optdet}

\begin{figure}[ht!]
\begin{center}
\includegraphics[width=\textwidth]{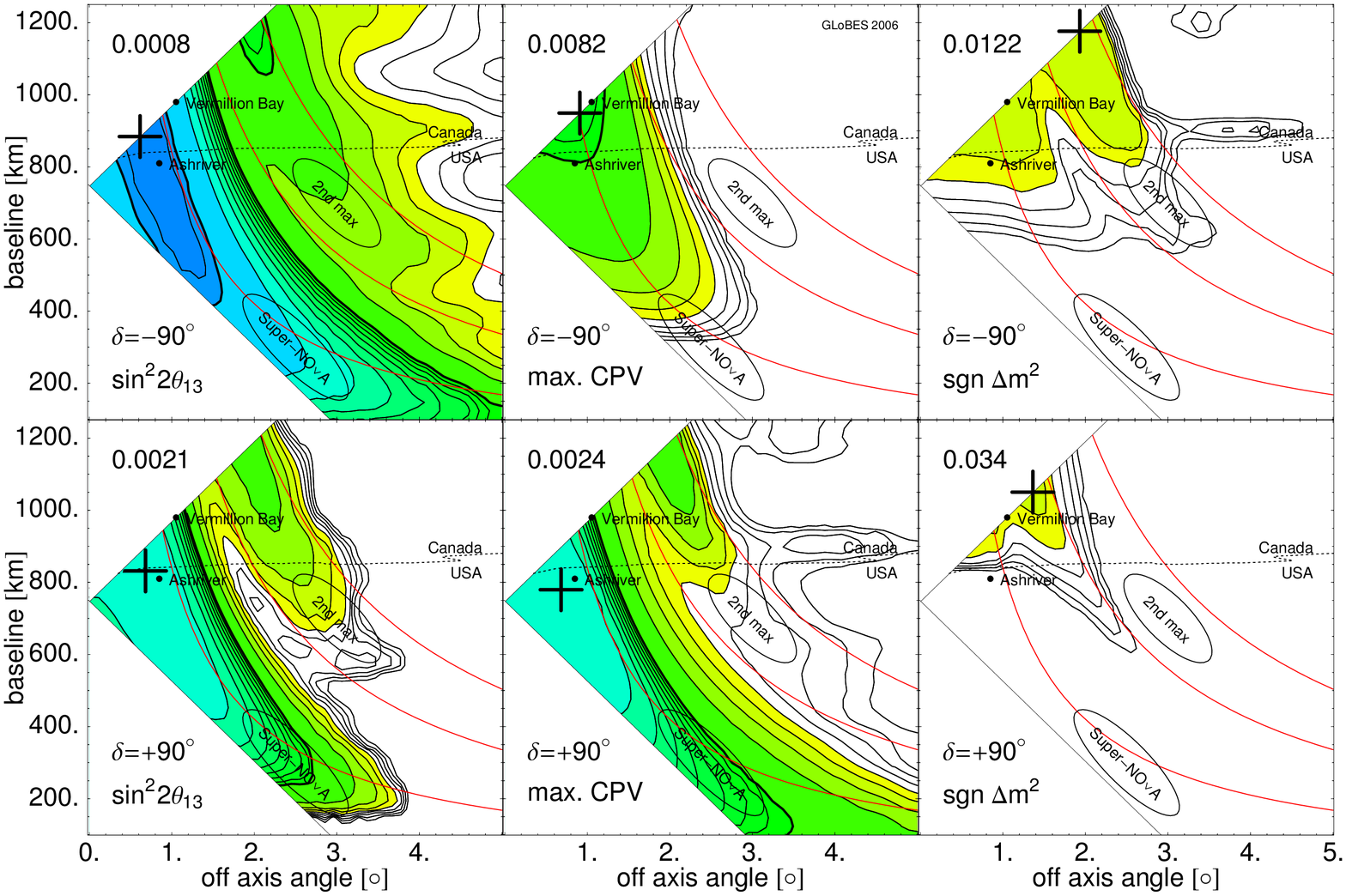}
\end{center}
\mycaption{\label{fig:leopt} Optimization of the NO$\nu$A* detector
  location (including phase~I at Ash River) in the off-axis
  angle--baseline plane. The plots show the $3\sigma$ discovery reaches 
   for nonzero $\stheta$, maximal CP violation, and the normal
   mass hierarchy. The bold plus signs
mark the locations with the best sensitivity. These best sensitivities are given as
  numbers in the upper left corner of each panel.
  The thin black contours are logarithmically
  spaced in $\stheta$, {\it i.e.}  $1,2,\ldots,9\cdot10^{n}$, where
  $n$ ranges from $-4$ to $-2$. The color shading ranges from
  blue/dark (best) to green/light (worst); the bold contours
  correspond to $\stheta=10^{-3}$ and $10^{-2}$. The solid thin red curves
  are iso-$L/E$ lines for the first, second and third
  oscillation nodes.  The dots correspond to specific selected sites.
 The dashed curve denotes the U.S.-Canadian border. The region below this curve
  is within the U.S.}
\end{figure}

In \figu{leopt}, we show the result of the optimization of the
NO$\nu$A* detector location (with phase~I at Ash River) for a
relatively wide range of angles and baselines. For this plot we
computed the sensitivity on a grid of 45 values in the off-axis angle
from $0^\circ$ to $5^\circ$ and 45 values in the baseline from
$100\,\mathrm{km}$ to $1250\,\mathrm{km}$. The plots show a
``conservative'' $\log_{10} \left( \stheta \right)$ reach at the $3
\sigma$ C.~L. for nonzero $\stheta$ (left column), CP
violation (middle column), and normal mass hierarchy (right column)
discovery potentials. The bold plus signs mark the locations with the
best sensitivities. In some cases there is a gap in the sensitivity as a
function of $\stheta$, \ie, the $\Delta\chi^2$-function assumes the
value $9$ (which corresponds to the $3\sigma$ C.~L.) at more than one
value of $\stheta$.  In these cases, the conservative reach is defined
as the largest $\stheta$ for which $\Delta \chi^2=9$. The upper row
corresponds to (the simulated) $\deltacp=-90^\circ$, the lower row to
$\deltacp=+90^\circ$.

For $\deltacp=+90^\circ$ (lower row), the Ash River location for the
second detector is indeed
optimal for $\stheta$ and CP violation.
Relatively short baselines around $200-400 \, \mathrm{km}$ perform
reasonably well too. The
Super-NO$\nu$A setup is 
marked in the figure by an ellipse with a corresponding label. 
A detector at the
second oscillation maximum has poor sensitivity because of low event
rates. The general region where such a second maximum detector could
be located is marked by an ellipse taken
from Refs.~\cite{Ayres:2004js,Messier} with a corresponding label.
Note that our iso-$L/E$ line which corresponds to the second
oscillation node does not intersect the region delimited by the ellipse.  
For the mass hierarchy sensitivity,
however, a longer baseline is necessary.  For $\deltacp=-90^\circ$
(upper row), the conclusions are similar for the Ash River site, but
the Super-NO$\nu$A configuration performs worse for CP violation.
The reason for the poor mass hierarchy performance of all short
baseline sites for $\deltacp=+90^\circ$ is that the sign-degeneracy
problem is most severe for this value of the CP phase.
As far as the optimal off-axis angle is concerned, we 
assume that NC backgrounds can be identified with a
LArTPC.  Therefore, in principle, an on-axis operation
would be possible. However, since the intrinsic beam background is
larger there, the exact on-axis position is not preferred.

\subsection{Risk minimization}

Two questions arise from \figu{leopt}: 
\begin{enumerate}
\item Is it possible to find one site which is a reasonable compromise
  for both cases of the CP phase?
\item
 Is it possible to fix a site given the current uncertainty on the
precise value of $|\Delta m^2_{31}|$?
\end{enumerate}

Concerning the dependence of the optimal position on the true value of
$\deltacp$, we observe that for the discovery of nonzero $\stheta$, the optimal
point does not move very much and is very close to Ash River. For CP
violation, the difference in optimal positions is somewhat larger, but
nevertheless the two points are close. Only for the mass hierarchy does
the difference becomes considerable.

In order to address this issue in a more quantitative way, we define a
\emph{relative sensitivity reach} (RSR): For each of the six panels in
\figu{leopt}, we divide the sensitivity reach at each point by its
value at the best point (bold plus symbol), and thus RSR $\ge
1$.  The optimal value of RSR is unity.
Next, the maximal RSR for the cases $\deltacp=+90^\circ$ and
$\deltacp=-90^\circ$ for each point in the plane is determined.  This
``maximal RSR'' corresponds to the risk-minimized optimum over the two
different values of $\deltacp$.
To illustrate this procedure, suppose that the maximum of the RSRs at
a point is $1.2$. At this point, the sensitivity reach would be within
$20\%$ of its optimal value for either case of $\deltacp=\pm90^\circ$.
The next step is to find the point which has the best maximal RSR for
each measurement, \ie, the smallest maximal RSR value.  This point
corresponds to the optimal detector location given the risk
minimization over $\deltacp=\pm90^\circ$.

\begin{figure}[t!]
\begin{center}
\includegraphics[width=\textwidth]{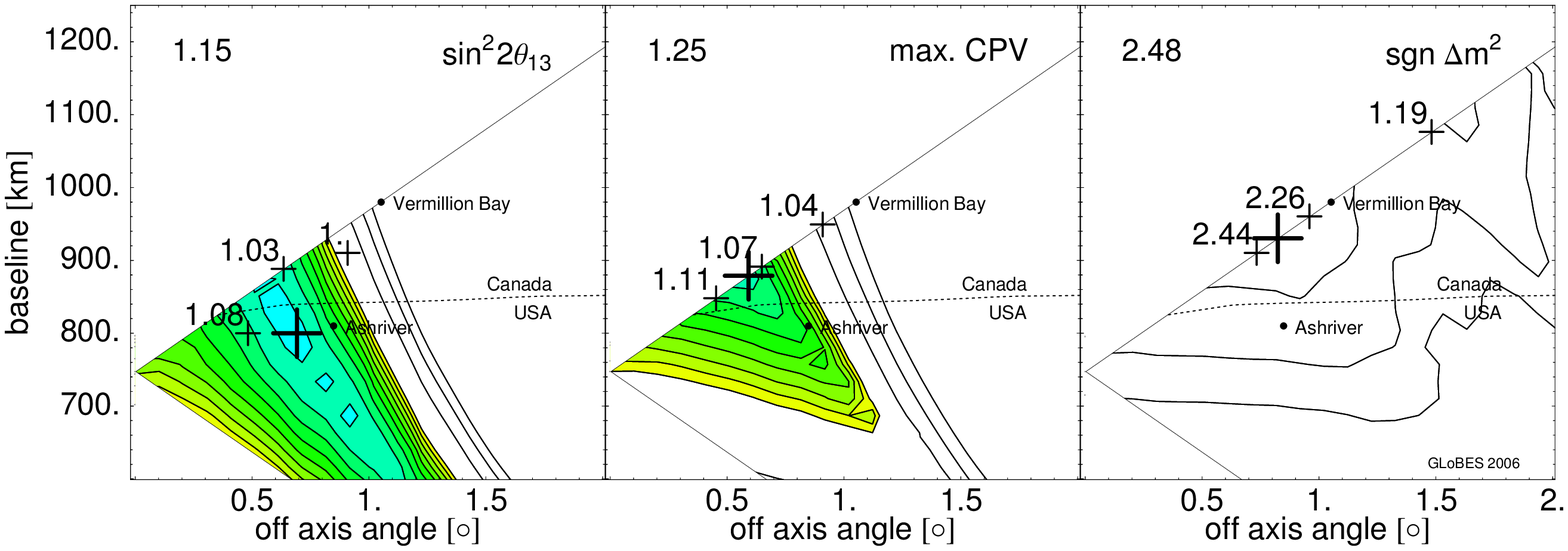}
\end{center}
\mycaption{\label{fig:optDM} Contours of constant maximum relative
  sensitivity reach (RSR) risk-minimized over $\deltacp=\pm90^\circ$
  and three values of $\Delta
  m^2_{31}=(2.0,2.5,3.0)\cdot10^{-3}\,\mathrm{eV}^2$.
  The shading ranges from $1.0$ (blue/dark) to $2.0$ (green/light) and
  the contours within the shading are in RSR steps of $0.1$. The
  contours in the unshaded area are in steps of $1.0$ extending up to
  $5$. The bold plus symbols denote the points where the smallest
  values occur. The absolute optimum is given as the number in the
  upper left corner of each panel.
  The small pluses denote the points at which the optimum value of the
  maximum RSR occurs for one specific value of $\Delta m^2_{31}$ only.
  The leftmost (rightmost) plus sign corresponds to $\Delta
  m^2_{31}=3\cdot10^{-3}\, (2\cdot10^{-3})\,\mathrm{eV}^2$.  The
  number adjacent to each plus sign is the value at the optimum. The
  dashed curve denotes the U.S.-Canadian border. The region below the
  curve is within the U.S.}
\end{figure}

In order to discuss the dependence on the unknown $\ldm$, we show in
\figu{optDM} the optimal detector locations for three different
choices of $\Delta m^2_{31}$ as the small plus symbols.  Let us first
 focus on these small plus symbols.  The leftmost pluses in each panel are for $\Delta
m^2_{31}=3.0\cdot10^{-3}\,\mathrm{eV}^2$, the middle ones are for our
standard value $\Delta m^2_{31}=2.5\cdot10^{-3}\,\mathrm{eV}^2$, and
the rightmost ones are for $\Delta
m^2_{31}=2.0\cdot10^{-3}\,\mathrm{eV}^2$. The numbers next to each
small plus symbol indicate the value of the maximal RSR at the points.
We observe that for the measurement of $\stheta$ (left panel) and CP
violation (middle panel) the maximal RSR never gets larger than $1.11$
for any choice of $\Delta m^2_{31}$. Moreover, the optima are not very
sensitive to $\Delta m^2_{31}$. For the mass hierarchy (right panel),
however the situation is less favorable. The optimal points are far
apart, and for a higher value of $\Delta m^2_{31}$ it is impossible to
find a location that has good sensitivity for both cases of
$\deltacp=\pm90^\circ$; the maximum RSR is $2.44$.

We then risk-minimize over $\deltacp$ and $\Delta
m^2_{31}$.  The contours shown in \figu{optDM} are contours of
constant RSR risk-minimized over the two values of $\deltacp$ and the
three values of $\Delta m^2_{31}$, \ie, the RSR value is maximized.
Therefore, these help to find the optimum location for each measurement
which takes into account our limited knowledge of the oscillation
parameters. The bold plus symbols denote the risk-minimized detector
locations. The maximal RSR values there are given in the upper left
corner of each panel. Again, for the discovery of $\stheta$ and CP
violation, we find that the values can be as low as $1.25$ and that
the two optima are not very far apart. This suggests that a joint
optimization for these two measurements is indeed possible, even if
one takes into account the large variation in $\Delta m^2_{31}$.  The
joint optimization results in a baseline $L=880\,\mathrm{km}$ at an
off-axis angle $0.6^\circ$. For the mass hierarchy, the situation is
more difficult, but clearly the main issue is not so much the
uncertainty in $|\Delta m^2_{31}|$ (which will be considerably reduced
by MINOS and NO$\nu$A phase~I), but the fact that the CP phase will
most likely be unknown at the time of the experiment. In this case,
the optimal location is at $L=930\,\mathrm{km}$ and an off-axis angle
$0.8^\circ$.

Further details on the optimization of the location of the NO$\nu$A*
detector are provided in the Appendix.


For all setups considered here we have assumed an equal fraction of
neutrino and antineutrino running. We have checked that this provides
superior sensitivities for all measurements. 
Specifically, our setup performs better than a pure
neutrino running option in combination with a short baseline detector,
such as considered in Refs.~\cite{MenaRequejo:2005hn,Mena:2005ri}.
We also
checked that the medium energy beam option does not provide better
sensitivities and would not significantly alter our conclusions.

In summary, we find that for the mass hierarchy measurement, a longer
baseline is preferable, whereas for $\stheta$ and CP violation the Ash
River site does very well. In addition, it is not far from
the optimum for the mass hierarchy. Since we do not find a
satisfactory optimum for all performance indicators simultaneously, we
choose Ash River for phase~II; see \Tab~\ref{tab:setups}.  Note that
we will not use phase~I for the comparisons (as for all of the other
experiments). Since we have the same site for phase~I and phase~II and
the luminosity for phase~II is much higher, it does not significantly
contribute to the sensitivities.

\section{Experiments with Wide Band Beams}
\label{sec:wbb}

Wide band beams are attractive because of their broad energy spectrum and higher 
on-axis flux~\cite{Beavis:2002ye,Diwan:2003bp}.
A potential limitation of using water Cherenkov detectors with wide-band beams was thought
to be the difficulty of isolating
QE events from the NC background from the high-energy tail of the spectrum. Recent work using
 ``Polfit'' in conjunction with several other discriminants, has shown that adequate
suppression of the $\pi^0$ background can be
achieved~\cite{Yanagisawa}. 
Motivated by this significant development, a comprehensive analysis of an
experiment with a 300 kT water Cherenkov
detector was carried out in Ref.~\cite{Barger:2006vy}.
A LArTPC may be a 
promising alternative to reduce backgrounds very efficiently while having
a higher efficiency through the utilization of non-QE events.

Currently, the most likely host for a wide band beam in the U.S. is
Fermilab. This leads to a number of interesting optimization issues.
First, since the Main Injector (MI) can provide protons with energies between
$\sim \, 30 \, \mathrm{GeV}$ and $120 \, \mathrm{GeV}$, where the
maximum efficiency is reached at around $60 \, \mathrm{GeV}$, the proton energy needs to be optimized. 
Second, the proposed baseline from Fermilab to a Deep Underground
Science and Engineering Laboratory (DUSEL) 
at the Homestake or Henderson mines leads to additional constraints
for the decay pipe since the boundaries of Fermilab limit the decay pipe
length. Compared to the NuMI beam which is pointed northerly
and with a $\sim 750 \, \mathrm{m}$ decay tunnel, 
the Fermilab-Henderson southwesterly baseline would constrain
the decay tunnel to less than $400 \, \mathrm{m}$. Therefore, one expects
a different beam spectrum for these two cases. Third, choosing between the
water Cherenkov and liquid argon detector technologies is
an interesting issue. We will discuss these main optimization topics in \Sec~\ref{sec:optwbb}.
Note that, compared to the last section, we do not use a specific beam line
and allow for arbitrary baseline--off-axis angle combinations. However, we
work within the length constraints for a southwesterly directed decay tunnel.
In some cases, we show the results for a northerly
directed tunnel for comparison. 

\subsection{Simulation Details}

\begin{table}[t]
\begin{center}
\begin{tabular}{llllcc}
\hline
Setup & Proton & Detector technology & $m_{\mathrm{Det}}$ & \multicolumn{2}{c}{Decay tunnel} \\
 & energy & & & Length & Direction \\
\hline
WBB-28$_\mathrm{S}$-WC & 28~GeV & Water Cherenkov & 300~kt & Short & Southwesterly \\
WBB-28$_\mathrm{S}$ & 28~GeV & LArTPC & 100~kt & Short & Southwesterly \\
WBB-60$_\mathrm{S}$ & 60~GeV & LArTPC & 100~kt & Short & Southwesterly \\
WBB-120$_\mathrm{S}$ & 120~GeV & LArTPC & 100~kt & Short & Southwesterly \\
WBB-120$_\mathrm{L}$ & 120~GeV & LArTPC & 100~kt & Long & Northerly \\
\hline
\end{tabular}
\end{center}
\mycaption{\label{tab:setupswbb} Different wide band beam setups used in this section. The 
subscripts ``$\mathrm{S}$'' and ``$\mathrm{L}$'' indicate whether the decay tunnel is shorter or 
longer than 400 m.
}
\end{table}

Our simulation is based on \Ref~\cite{Barger:2006vy}. 
We use five years of neutrino running at 1 MW target power, and five years
of antineutrino running at 2 MW target power. The baseline is $1290 \, \mathrm{km}$
corresponding to Fermilab-Homestake. As a detector, we either use a $300 \, \mathrm{kt}$ water
Cherenkov detector, or a $100 \, \mathrm{kt}$ liquid argon TPC.
Details of the LArTPC simulation are given in \Sec~\ref{sec:novadetails},
while details of the water Cherenkov detector simulation can be found 
in \Ref~\cite{Barger:2006vy}. We use a systematic uncertainty of 5\% on both
signal and background.
All of our setups are designed for the Fermilab MI as the proton source.
As proton energies, we use $28 \, \mathrm{GeV}$ corresponding to the old BNL proposal,
$60 \, \mathrm{GeV}$ corresponding to the start of the maximum efficiency region, and
$120 \, \mathrm{GeV}$ corresponding to the current proton energy of the MI.
All of these setups use the southwesterly pointed short decay tunnel. In addition,
we show the results for $120 \, \mathrm{GeV}$ protons and a new long northerly directed 
decay tunnel similar to the NuMI tunnel. 
For all setups, we use a LArTPC, but we compare it
to the water Cherenkov detector for the $28 \, \mathrm{GeV}$ protons. Our setups
are summarized in \Tab~\ref{tab:setupswbb}.


\subsection{Motivation for On-Axis Operation}

\begin{figure}[t!]
\begin{center}
\includegraphics[width=\textwidth]{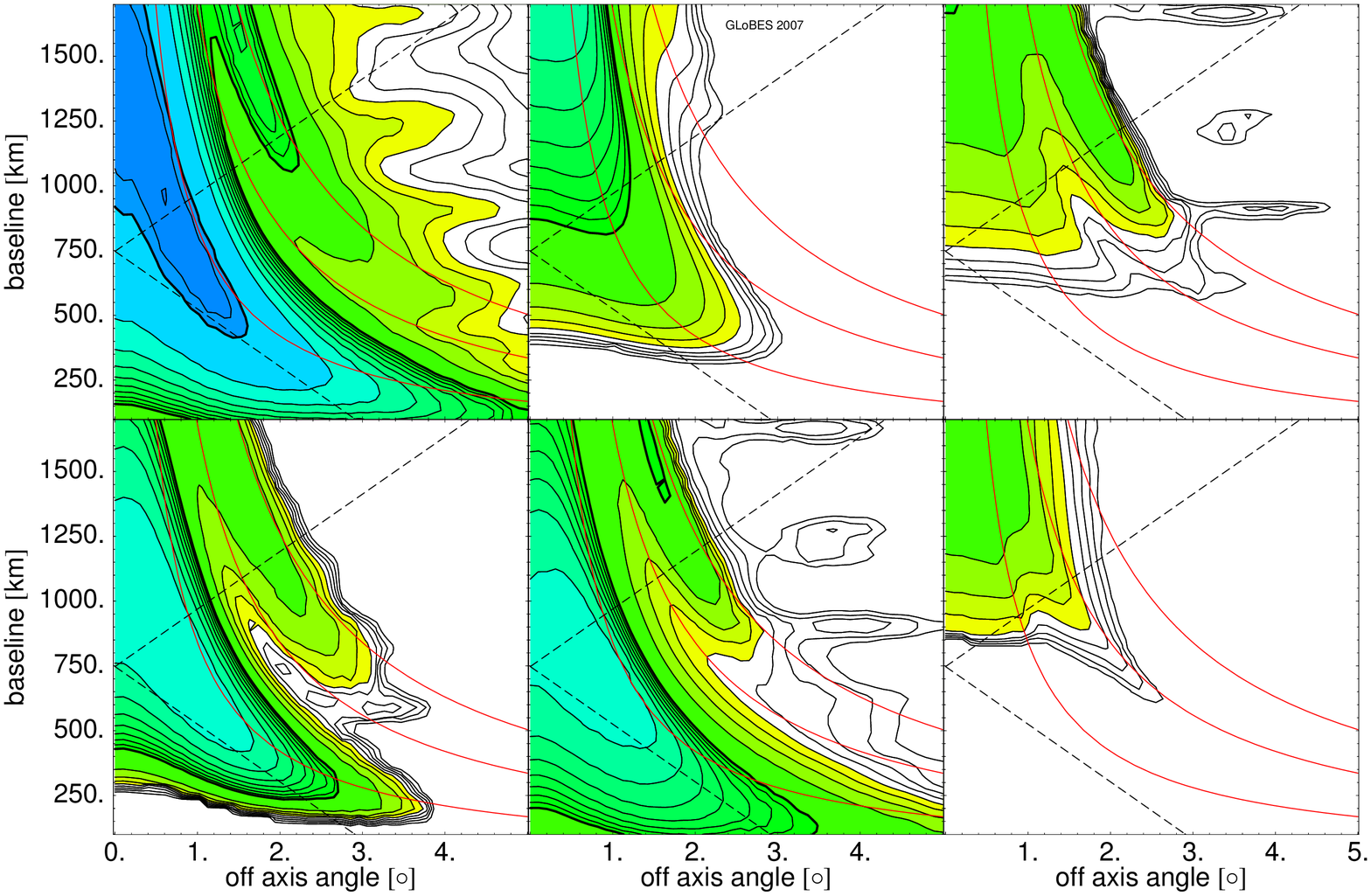}
\end{center}
\mycaption{\label{fig:leoptall} Optimization for a hypothetical NO$\nu$A* experiment 
(including phase~I at Ash River) in the off-axis
  angle--baseline plane, but without being constrained by the NuMI beam line. 
Unlike \figu{leopt}, the full ranges for baselines and off-axis angles are shown.
As in \figu{leopt}, the plots show the discovery reaches at the $3\sigma$ C.~L. 
   for nonzero $\stheta$, maximal CP violation, and the normal
   mass hierarchy. 
  The thin black contours are logarithmically
  spaced in $\stheta$, {\it i.e.,}  $1,2,\ldots,9\cdot10^{n}$, where
  $n$ ranges from $-4$ to $-2$. The color shading ranges from
  blue/dark (best) to green/light (worst); the bold contours
  correspond to $\stheta=10^{-3}$ and $10^{-2}$. The solid thin red curves
  are iso-$L/E$ lines for the first, second and third
  oscillation nodes. The constraints from the
NuMI decay tunnel are shown as dashed lines for comparison.}
\end{figure}

Before we address the main optimization issues for a WBB experiment, 
let us motivate the choice of baseline
and off-axis angle, and compare the WBB to NO$\nu$A*. 
In \figu{leoptall}, we show the performance for 
 a hypothetical NO$\nu$A* experiment without constraints from the NuMI beam line, 
but for the same beam. \ie, we have allowed for the possibility of a new decay tunnel. 
The detector parameters and running times are those in \Sec~\ref{sec:nova}. 
Note that detector locations within the dashed lines can be
accommodated within the NuMI beam constraints.
Unlike \figu{leopt}, the full ranges for baselines and off-axis angles are shown. 

As can be seen from this figure, the choice of $L \simeq 1200-1500 \, \mathrm{km}$ 
performs well for all measurements, and in (almost) all cases at least as well 
as the best possible location with the NuMI decay tunnel constraint. 
In addition, there is no need to go off-axis,
because the NC rejection is assumed to be very efficient for the LArTPC.
Note that the on-axis performance of even longer baselines has been studied in  
\Ref~\cite{Barger:2006vy}, and baselines between 
about $1200-1500 \, \mathrm{km}$ were found to be optimal.
Similarly, we find that the WBB on-axis concept with a baseline of $\sim 1290 \, \mathrm{km}$
in combination with a LArTPC is close to optimal for all performance indicators.
Note that the NuMI-like beam line and beam correspond to our wide band beam option 
WBB-120$_\mathrm{L}$, \ie,
the northerly directed decay tunnel. 
Compared to NO$\nu$A*, the detector is on-axis and at a longer baseline. 

\subsection{Optimization of a Fermilab-Based Wide Band Beam Experiment}
\label{sec:optwbb}

\begin{figure}[t!]
\begin{center}
\includegraphics[width=\textwidth]{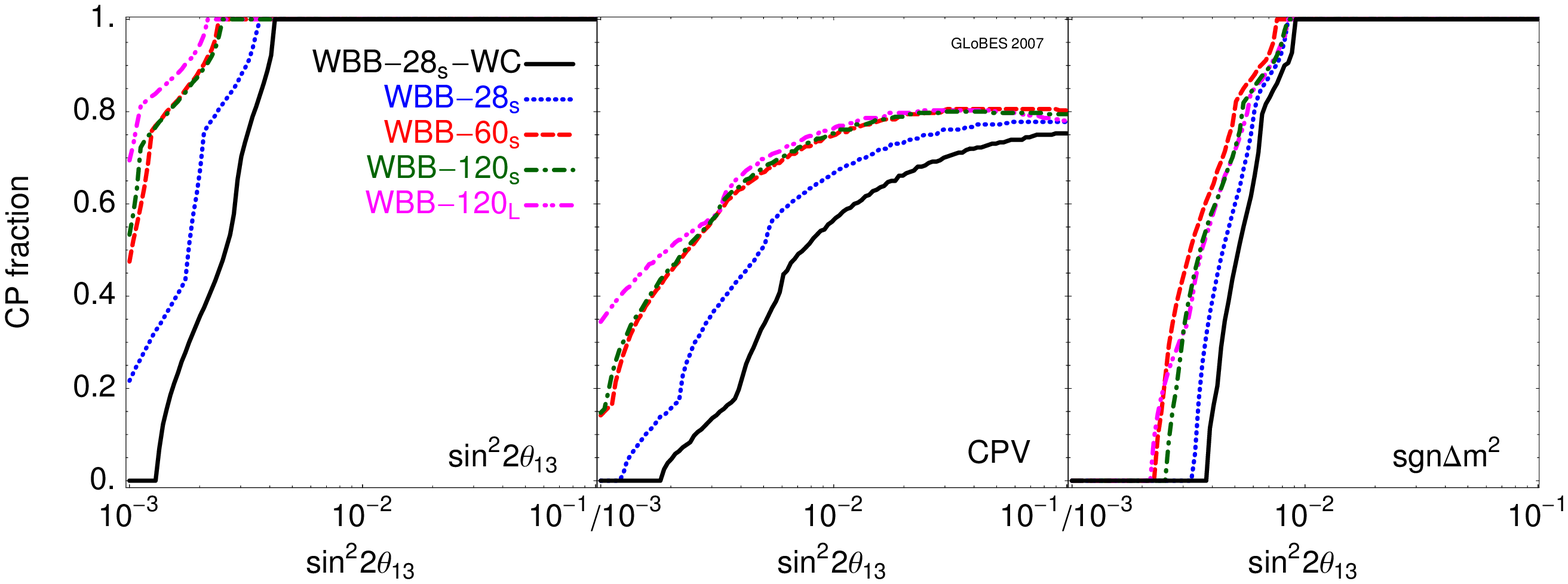}
\end{center}
\mycaption{\label{fig:wbbcomp} The $3\sigma$ discovery potentials for a 
nonzero $\stheta$, CP violation, and
normal mass hierarchy of the WBB options 
in \Tab~\ref{tab:setupswbb}.
}
\end{figure}

In order to discuss the impact of proton energy, decay pipe length, and detector technology,
in \figu{wbbcomp}, we show the $\stheta$, CP violation, and
normal mass hierarchy discovery potentials of the WBB options in \Tab~\ref{tab:setupswbb}.
The performance of the setups improves as one moves down in \Tab~\ref{tab:setupswbb}.
Note that for the option with a water Cherenkov detector, 
the exposure is a factor of three higher than the options with a LArTPC. 

As far as the proton energy is concerned, compare the options WBB-28$_\mathrm{S}$, 
WBB-60$_\mathrm{S}$, and WBB-120$_\mathrm{S}$, which are all for the same decay pipe 
length but for different proton energies. The performance improves with 
proton energy, but the  WBB-60$_\mathrm{S}$ setup is already close to optimum because 
the maximum efficiency (of the whole system) is reached for 60 GeV protons. 
A comparison of WBB-120$_\mathrm{S}$ and WBB-120$_\mathrm{L}$ indicates that the beam with 
the longer decay pipe performs better for small $\stheta$, which means that a NuMI-like beam is 
 the best choice. However, Fermilab boundaries
prohibit such a long decay pipe for the considered Fermilab-Homestake or -Henderson options, 
making  WBB-120$_\mathrm{S}$ the viable option. Note that a somewhat lower 
proton energy hardly has any impact on the 
physics performance, whereas a longer decay pipe could definitively help. 
To assess detector technologies, compare the
WBB-28$_\mathrm{S}$ (liquid argon) and WBB-28$_\mathrm{S}$-WC setups. 
For the chosen detector masses of $100 \, \mathrm{kt}$ for liquid argon and  
$300 \, \mathrm{kt}$  for the water Cherenkov detector, respectively, the liquid argon setup 
performs significantly better. 
We find that 4 kt 
of water are equivalent to one kt of liquid argon for the $\stheta$ discovery, 4.4 kt of water 
are equivalent to one kt of liquid argon for the CP violation discovery, and 3.6 kt 
of water are equivalent to one kt of liquid argon for the mass hierarchy discovery. 
This means that (at least for the $28 \, \mathrm{GeV}$ proton energy) the cost per kt 
of water has to be less than 25\% of the cost per kt of liquid argon in order to choose 
water as the detector material. Note that this ratio can not be easily extrapolated to higher 
proton energies because of the higher neutrino energies and therefore higher NC background. 


For the following physics comparison, we choose WBB-120$_\mathrm{S}$. 

\section{Physics Comparison}

In this section, we compare the chosen setups from each of the previous sections. 
In addition, we include a neutrino factory (NuFact) and a $\beta$-beam setup in some of the 
discussion. Our setups for this section are summarized in \Tab~\ref{tab:setups}. 

\begin{table}
{\small
\begin{tabular}{lccccccr}
\hline
Setup & $t_{\nu}$ [yr] &  $t_{\bar{\nu}}$
[yr] & $P_\mathrm{Target}$ [MW] & $L$ [km] & Detector technology &
$m_{\mathrm{Det}}$ [kt] & $\mathcal{L}$ \\ 
\hline
NO$\nu$A* & 3 & 3 & 1.13 ($\nu$/$\bar{\nu}$) & 810 &
LArTPC & 100 & 1.15 \\
WBB-120$_\mathrm{S}$ &  5 &  5 & 1 ($\nu$) +2
($\bar{\nu})$& 1290 & LArTPC & 100 & 2.55 \\
T2KK &  4 & 4 & 4 ($\nu$/$\bar{\nu}$)&
295+1050 & Water Cherenkov & 270+270 & 17.28 \\
\hline
$\beta$-beam & 4 & 4 & n/a & 730 & Water Cherenkov & 500 & n/a \\
NuFact &  4 & 4 & 4 & 3000+7500 & Magn. iron calor. & 50+50 & n/a
\\
\hline
\end{tabular}
} 
\mycaption{\label{tab:setups} Setups considered, neutrino $t_{\nu}$ and antineutrino 
$t_{\bar{\nu}}$ running times, corresponding target power
$P_\mathrm{Target}$, baseline $L$, detector technology, detector mass
$m_{\mathrm{Det}}$, and exposure $\mathcal{L}$ [$\mathrm{Mt \, MW \, 10^7 \, s}$]. 
For beta beam and NuFact,
we assume $10^7 \, \mathrm{s}$ of operation/year. Target power does
not apply to an ion source used for the beta beam. 
We use $2.9\cdot10^{18}$ useful $^6$He decays/year and
$1.1\cdot 10^{18}$ useful $^{18}$Ne decays/year for the beta beam, 
where we have $\gamma=350$ for both polarities.
Details on the simulation can be found in \Ref~\cite{Burguet-Castell:2005pa}. 
For the neutrino factory, we use $10^{21}$ useful muon decays/year for both polarities, and 
two magnetized iron calorimeters at two different
baselines. For details, see \Refs~\cite{Huber:2002mx,Huber:2003ak,Huber:2006wb}.}
\end{table}

\subsection{Exposure Scaling and Normalized Comparison}

\begin{figure}[tp!]
\begin{center}
\includegraphics[height=18cm]{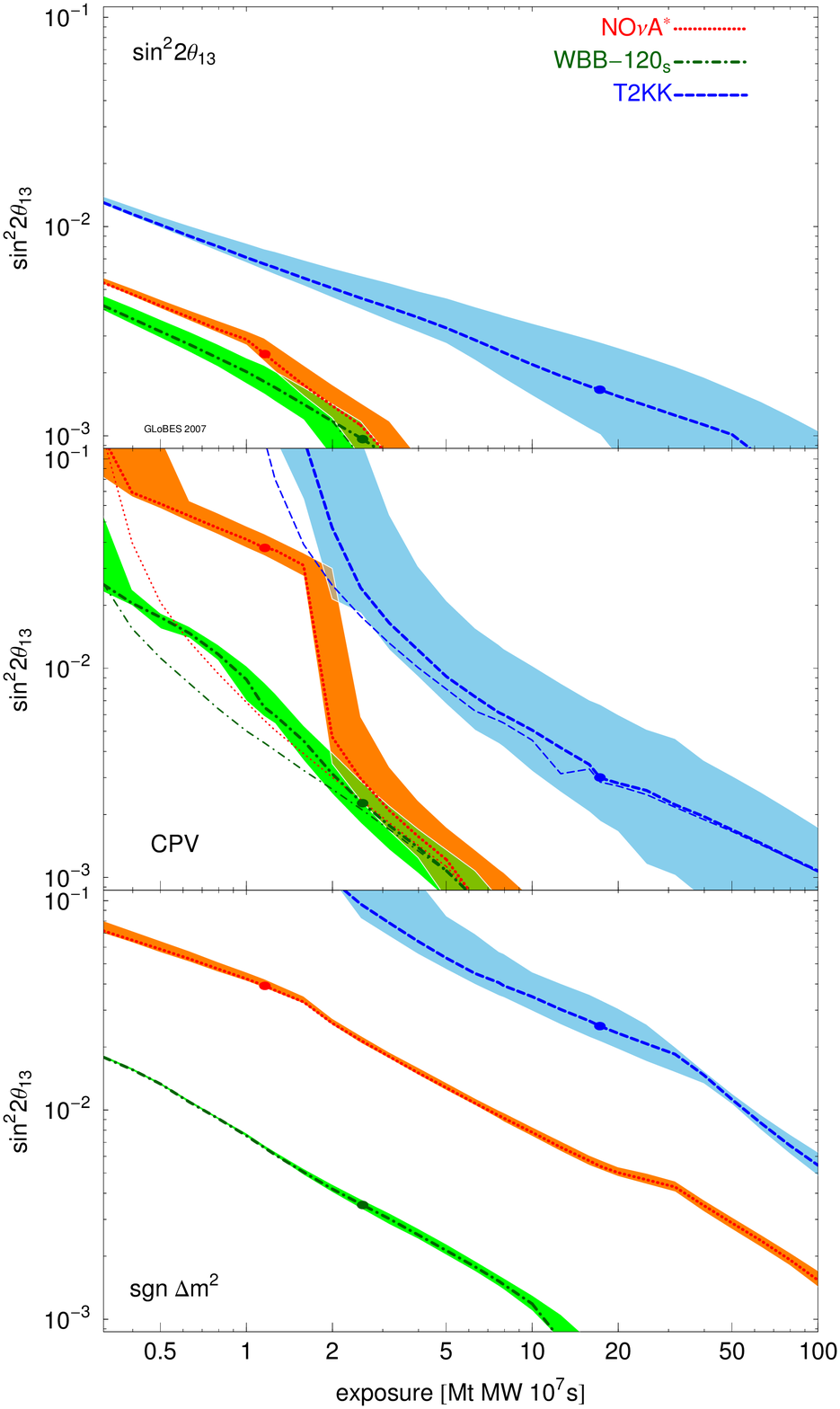}
\end{center}
\mycaption{\label{fig:lumiscale} The $\stheta$ reach at $3\sigma$ 
for the discovery of nonzero $\stheta$, CP violation, and the normal hierarchy as a 
function of exposure. The curves are for 
a fraction of $\deltacp$ of 0.5, 
which means that the performance will be better for 50\% of all 
values of $\deltacp$, and worse for the other 50\%. 
The light curves in the CPV panel are made under the assumption that the mass hierarchy 
is known to be normal. 
The dots mark the exposures of the setups as defined 
in \Tab~\ref{tab:setups}. The shaded regions result by varying the systematic uncertainties from 
2\% (lower edge) to 10\% (upper edge). 
}
\end{figure}

In order to compare the normalized performance of different experiments, in \figu{lumiscale}
we show the $3\sigma$ reaches for the discovery of nonzero $\stheta$, CP violation, and the 
normal hierarchy as functions of exposure. The dots mark the nominal exposures of 
the setups as given in \Tab~\ref{tab:setups}. The shaded regions show the dependence of the
sensitivities on the level of systematic uncertainties. The lower (upper) edge of each region
 corresponds to a systematic uncertainty of 2\% (10\%). Observe that the detection of CP
violation is more sensitive to the systematic uncertainty than the other measurements. 

Most of the discovery reaches in \figu{lumiscale} scale more or less like statistics, 
where there may be a slight transition from rate to spectrum dominated regimes 
(see, for instance, CPV for T2KK). The only severe exception is the NO$\nu$A* scaling of 
the CPV reach, which was noticed and explained in Ref.~\cite{Barger:2006kp}. 
In this case, a slightly higher luminosity can have a tremendous impact 
by the resolution of the mass hierarchy degeneracy (\cf, light curves for known mass hierarchy).
 A factor of two higher luminosity could increase the $\stheta$ reach by an order of magnitude. 

\begin{figure}[t!]
\begin{center}
\includegraphics[width=\textwidth]{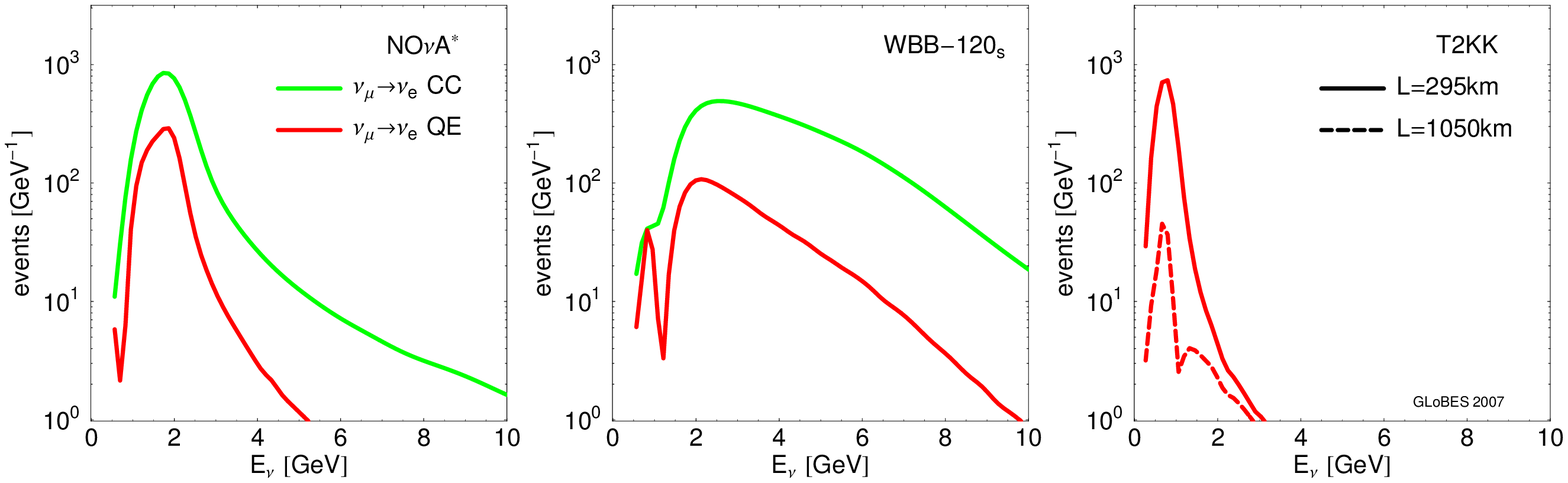}
\end{center}
\mycaption{\label{fig:rates} Neutrino event rate spectra for $\stheta=0.04$, $\deltacp=0$
for NO$\nu$A*, WBB-120$_\mathrm{S}$ and T2KK. All spectra are computed for an exposure of 
$1 \, \mathrm{Mt \, MW \, 10^7 \, s}$ including detection efficiencies and energy resolution. 
For NO$\nu$A* and WBB, red/dark curves are the QE events only, and the
green/light curves are the non-QE events only. For T2KK, the solid curve refers to the 
detector in Japan, the dashed
curve to the detector in Korea.
}
\end{figure}

NO$\nu$A* and WBB-120$_\mathrm{S}$ are approximately equal concepts for large 
enough exposures $\gtrsim 2$ $\mathrm{Mt \, MW \, 10^7 \, s}$ for the $\stheta$ and 
CP violation discoveries. However, for the mass hierarchy discovery, WBB-120$_\mathrm{S}$ 
performs better because of the longer baseline. 
The curves for the T2KK concept are all above the ones for the Fermilab-based setups. 
In this case, one has to take into account the lower cost of water compared to liquid argon 
as detector material.  However, the lower neutrino energies and the
low event rates in the detector in Korea, highly affect the
competitiveness. One can also see these properties in the event rate spectra for the 
same exposure in \figu{rates}. The NO$\nu$A* and WBB-120$_\mathrm{S}$ options have 
broader spectra that peak at higher energies, and the integrated event rate
is much higher.

\subsection{Prediction for Nominal Exposure and its Robustness}

\begin{figure}[t!]
\begin{center}
\includegraphics[width=\textwidth]{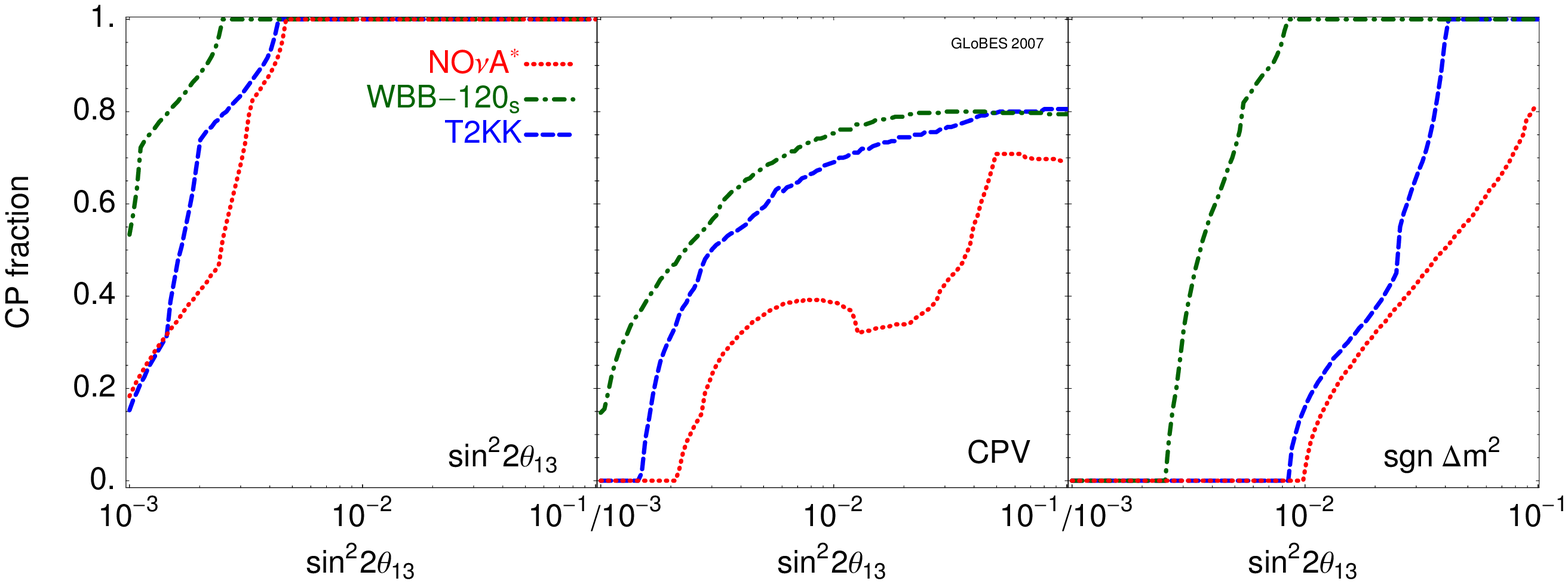}
\end{center}
\mycaption{\label{fig:sumnominal}  Comparison of superbeam upgrades in the
configurations of Table~\ref{tab:setups} at the $3 \sigma$ C.~L. The plots show the
discovery reaches
for a nonzero $\stheta$, CP violation,
and the normal hierarchy.
}
\end{figure}

We compare in \figu{sumnominal} the absolute performance for the exposures given 
in \Tab~\ref{tab:setups}. 
For all performance indicators, the WBB-120$_\mathrm{S}$ has the strongest physics potential 
and NO$\nu$A* the weakest. All experiments can discover $\stheta$, CP violation, and 
the mass hierarchy for large $\stheta$ for most values of $\deltacp$, but there are 
substantial differences for smaller values of $\stheta$. As we have seen above, 
this performance comparison changes on normalizing the exposure.

\begin{figure}[t!]
\begin{center}
\includegraphics[width=\textwidth]{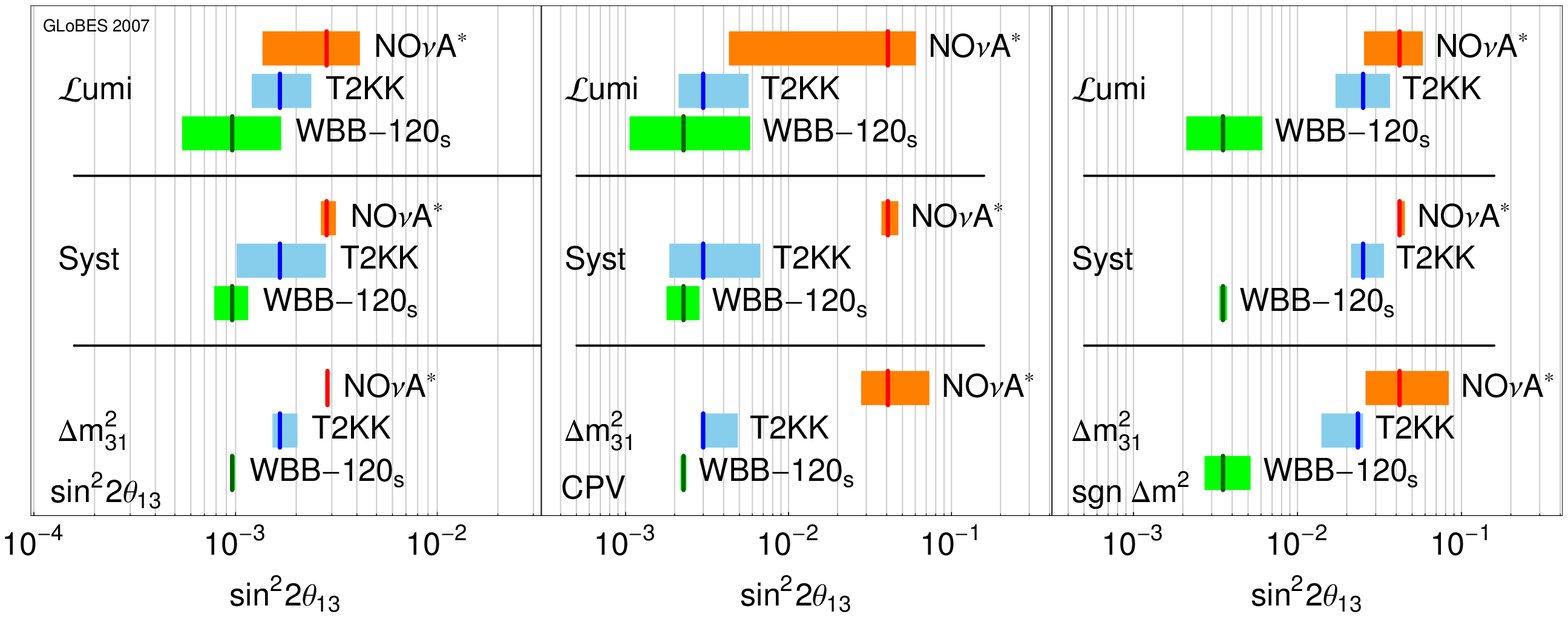}
\end{center}
\mycaption{\label{fig:summarybars} Robustness with respect to the main impact factors 
luminosity/exposure, systematics prediction, and true $\ldm$. We show the impact on the 
$3\sigma$ reaches for a nonzero $\stheta$, CP violation, and the mass hierarchy, assuming
a CP fraction of 0.5. The bold vertical lines within the bars correspond 
to our standard assumptions. The bars mark the ranges between the best case assumptions 
(double exposure, 2\% systematic errors, $\ldm=3.0 \, 10^{-3} \, \mathrm{eV}^2$) 
and the worst case assumptions (half exposure, 10\% systematic errors, 
$\ldm=2.0 \, 10^{-3} \, \mathrm{eV}^2$) for the respective impact factor. 
}
\end{figure}

Let us assess how robust the conclusions based on the nominal exposure are under the 
following three main impact factors which could affect the performance of a superbeam upgrade:
\begin{enumerate}
\item 
 The originally anticipated luminosity cannot be reached, or an improvement of the 
proton plan leads to a substantial increase of the originally anticipated luminosity. 
To account for this possibility, 
we vary the exposure between half and twice the exposure we have used so far.
\item
 The original sytematic error estimate turns out to be too optimistic or too conservative. 
We vary the signal and background normalization errors between 2\% to 10\%.
We have assumed a 5\% uncertainty throughout.
\item
 The current best-fit value of $\ldm$ the experiments are optimized for turns out to be 
somewhat different from the actual value. We vary $\ldm$ between 
$2.0 \, 10^{-3} \, \mathrm{eV}^2$and $3.0 \, 10^{-3} \, \mathrm{eV}^2$.
\end{enumerate}
The results of this robustness analysis are displayed in \figu{summarybars}. 
Each bar reflects the range from a too conservative original estimate (left end) 
to the original assumption (bold vertical lines) to a too optimistic original estimate 
(right end).

From the upper row of the CPV panel, we see that a factor of two higher luminosity 
for NO$\nu$A* leads to a much better CP violation performance. 
The impact of an incorrect systematics estimate is strongest on T2KK and weakest on NO$\nu$A*.
Experiments using a LArTPC are much
less sensitive to variations of the systematic errors since the
background levels are much lower. This also implies that T2KK has the most to gain 
from a better understanding of systematics. 
On the other hand, increasing the systematics from 5\% to 10\% in T2KK would
require to increase the exposure by nearly a factor three to compensate the loss of sensitivity 
to nonzero $\stheta$ and $\deltacp$. 
The sensitivity of experiments with narrow band beams,  
NO$\nu$A* and T2KK, is very much affected by the actual value of $\ldm$, 
whereas the sensitivity of 
WBB-120$_\mathrm{S}$ is hardly affected. Note that in this case, we observe the 
strongest impact on NO$\nu$A*.


The ranges covered by the bars provide a measure of risk. 
 Since the bars for the $\stheta$ and CP violation 
discoveries overlap for the different setups, they may yield very similar results 
given their nominal exposures.
Only for the mass hierarchy measurement, WBB-120$_\mathrm{S}$ is significantly better 
than the other two experiments. 

\begin{figure}[t!]
\begin{center}
\includegraphics[width=\textwidth]{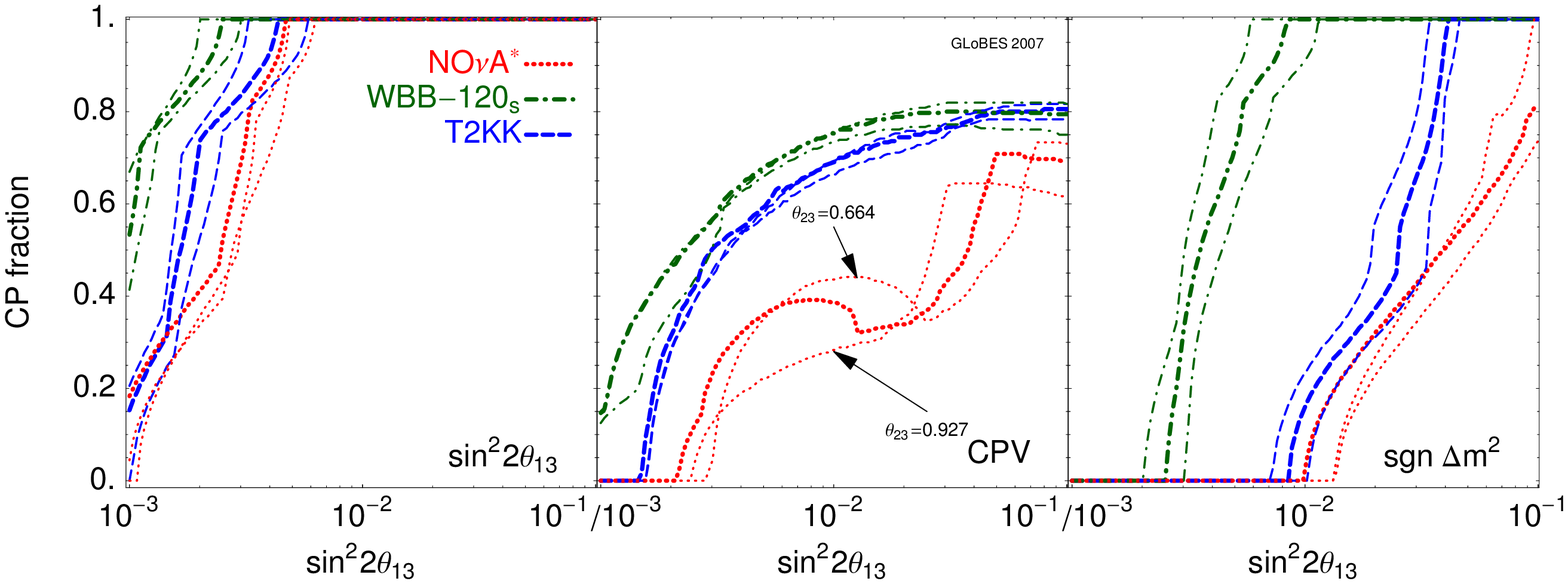}
\end{center}
\mycaption{\label{fig:octant} Effect of the octant degeneracy on the $3\sigma$ 
sensitivities of the experiments. The three sets of curves for each experiment correspond to
$\theta_{23}=0.664$ , $\pi/4$ (thick lines) and $0.927$. Other than for NO$\nu$A* (for
which a degenerate solution limits the sensitivities), the naive
expectation that the sensitivities are greater for larger 
$\theta_{23}$, is borne out. }
\end{figure}

Another important consideration is the effect of the octant degeneracy which appears 
for deviations of $\theta_{23}$ from $\pi/4$. 
In Fig.~\ref{fig:octant}, we show the sensitivities of each setup under 
the assumption that $\theta_{23}=0.664$ or $0.927$,
which are at the extremes of the 2$\sigma$ interval allowed by atmospheric data. 
So long as $\theta_{23}$
does not deviate substantially from $\pi/4$, and become inconsistent with current data,
this degeneracy does not affect our results significantly. All experiments besides
NO$\nu$A* show a simple rescaling of their sensitivities. For NO$\nu$A*, the
wrong-octant, wrong-hierarchy solution impacts the CPV and the mass hierarchy
measurements.

Finally, we consider how assumptions about the energy resolution affect our results.
Spectral information
is relevant only when statistics are large, {\it i.e.}, when
$\theta_{13}$ is large. For small $\theta_{13}$ the energy resolution is of little relevance
since spectral information does not enhance the sensitivity.  The energy resolution
of the water Cherenkov detector is based on a simulation of the Super-Kamiokande 
detector~\cite{Yanagisawa}. 
For the options using a LArTPC, the energy
resolution is more uncertain. We explicitly checked that for $\sin^22\theta_{13}=0.1$,
varying the energy resolution for QE events between 5\%/$\sqrt E$ and 20\%/$\sqrt E$
barely affects the fraction of CP phases for which a given measurement can be
performed. The typical variation is less than 0.01. The only exception occurs for the
determination of the mass hierarchy with NO$\nu$A*, for which the CP fraction 
changes from 0.84 at 5\%/$\sqrt E$ to 0.72 at 20\%/$\sqrt E$.

\subsection{Comparison with a Neutrino Factory and a $\boldsymbol{\beta}$-Beam Experiment}

For small $\stheta \ll 10^{-2}$, it is well-known that a neutrino factory complex has the 
optimal physics potential for all of the considered performance indicators 
(\cf, \eg, \Ref~\cite{Huber:2006wb}). This is a consequence of 
the high neutrino energies and high event rates. However, the oscillation maximum 
sits at relatively low energies, where the backgrounds from event misidentification 
are large, and the event rates are comparatively moderate. Therefore, for 
large $\stheta$, a $\beta$-beam or superbeam experiment tuned to the oscillation 
maximum may have the better performance. Since the effort for a $\beta$-beam may 
be larger than for a superbeam upgrade, and the technology needs further exploration, 
it is an interesting question if the superbeam
upgrades can compete with a neutrino factory or $\beta$-beam for large $\stheta$.
We use the neutrino factory and $\beta$-beam setups from \Tab~\ref{tab:setups} for this 
comparison.\footnote{Note that some gain in the neutrino factory or $\beta$-beam performance 
can be obtained by a reoptimization for large $\stheta$ (see, \eg, 
\Refs~\cite{Huber:2005jk,Huber:2006wb,Geer:2007kn}). However, this gain is usually moderate.}

All the experiments under consideration have good sensitivity to nonzero $\stheta$
and the mass hierarchy for large $\stheta$.
We therefore do not discuss the $\stheta$ and mass hierarchy sensitivities
and focus on the CP violation measurement.

\begin{figure}[tp!]
\begin{center}
\includegraphics[height=18cm]{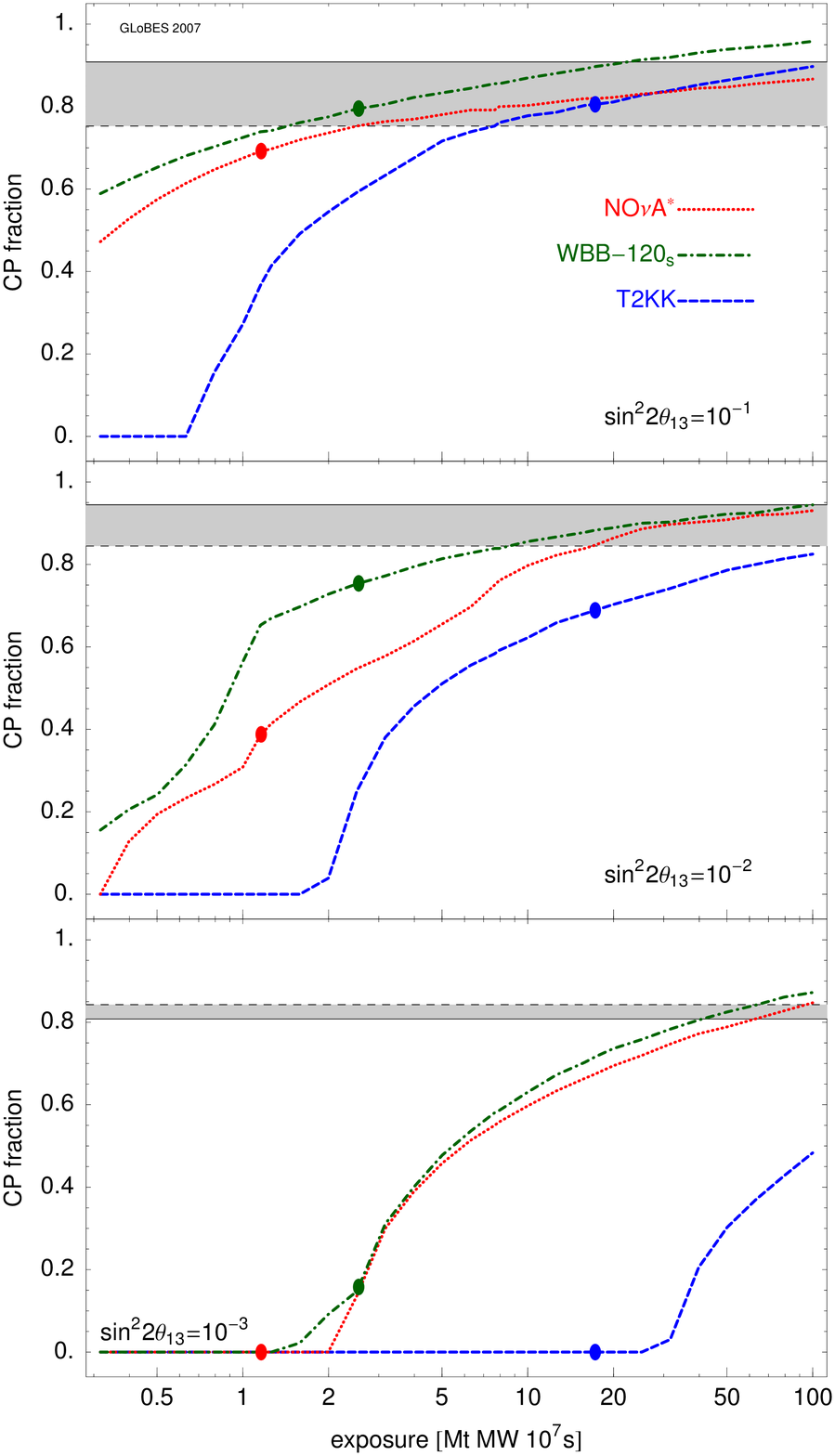}
\end{center}
\mycaption{\label{fig:largeth13} CP fraction for a $3\sigma$ discovery of CP violaton as a function of
exposure. The different panels correspond to $\stheta=0.1$,  $\stheta=0.01$ and  $\stheta=0.001$. 
The shaded region marks the potential between a $\beta$-beam (solid line) and neutrino factory 
(dashed line) as defined in \Tab~\ref{tab:setups}.
}
\end{figure}

In \figu{largeth13} we show the CP fraction for the $3\sigma$ discovery of CP violaton as a function of
exposure. The different panels correspond to different true values of $\stheta$. The shaded region 
marks the potential between a $\beta$-beam (solid line) and neutrino factory (dashed line) as 
given in \Tab~\ref{tab:setups}. 
For $\stheta=0.1$, the superbeam upgrades perform
at least as well as the neutrino factory, and a moderate increase of exposure can 
make their physics potential optimal. Note that, for instance, the neutrino factory 
requires a target power of $4 \, \mathrm{MW}$. The Fermilab-based experiments therefore still 
have space for an increase of target power.
For $\stheta=0.01$, the situation is already very different. In this
case, WBB-120$_\mathrm{S}$  can compete with the neutrino factory with a factor of 
two or three increase in the exposure. This
upgrade basically corresponds to an upgraded proton source and a somewhat longer running time. 
The increase in exposure necessary for NO$\nu$A* and T2KK to be competitive is unrealistic.
For $\stheta=0.001$, experiments with upgraded superbeams cease to be competitive.

The above discussion indicates that superbeams 
may be the technology of choice for large $\stheta$. However, if $\stheta$ is not discovered 
in the coming generation of
superbeam and reactor experiments, one will have to consider a neutrino factory or 
$\beta$-beam optimized
for a maximum reach in $\stheta$.

\section{Summary}

The important points of our paper that address the goals listed in the introduction are:

\begin{enumerate}
\item
 T2KK: Placing half the fiducial mass of the detector at 1050~km improves the sensitivity
to the mass hierarchy and to CP violation. Using identical detectors at the 1050~km and 295~km 
baselines so that the systematics are fully correlated only improves the sensitivity to the
mass hierarchy.
\item The optimal location for a NO$\nu$A* detector within the U.S. is the Ash River site.
Better risk-minimized mass hierarchy and CP violation
sensitivities can be obtained with a baseline $\sim 880 - 930$ km.
\item The optimal baseline for a wide band beam experiment
is between 1200 and 1500 km. High proton energies and a long decay tunnel are preferable.
\item  Among experiments with super neutrino beams, 
wide band beam experiments have the most robust performance and the 
best mass hierarchy performance. The sensitivity of experiments with narrow band beams 
is significantly affected by the true value of $|\Delta m^2_{31}|$. Overall,
 wide band beam experiments are the best experimental concept. 
\item  NO$\nu$A* is competitive with WBB-120$_\mathrm{S}$ for the discovery of nonzero 
$\theta_{13}$ and CP violation for exposures above $2 \, \mathrm{Mt \, MW \, 10^7 \, s}$.
\item Within the context of wide band beam experiments,
 approximately 4t of water performs as well as  1t of liquid argon. However, this ratio
may be different for proton energies higher than 28~GeV.
\item
 Superbeam versus $\beta$-beam/NuFact: WBB-120$_\mathrm{S}$ is a better experimental
concept than a neutrino
factory if $\stheta \gtrsim 0.01$. For $\stheta \lesssim 0.01$, experiments with 
superbeams are not competitive. For the exposures listed in \Tab~\ref{tab:setups}, if
\subitem
(a) $\stheta \gtrsim 0.04$: T2KK or WBB-120$_\mathrm{S}$ can make all three measurements with similar 
sensitivities as a $\beta$-beam/NuFact. 
\subitem
(b)  $0.01 \lesssim \stheta \lesssim 0.04$: Only WBB-120$_\mathrm{S}$ performs as well as a
$\beta$-beam/NuFact. T2KK is unable to establish the mass hierarchy with equivalent sensitivity.
\end{enumerate}

\newpage
\subsection*{Acknowledgments}

We thank E. Beier, M. Bishai, M. Dierckxsens, M. Diwan and B. Fleming for useful discussions.
This research was supported by the U.S.
Department of Energy under Grants No. DE-FG02-95ER40896 and
DE-FG02-04ER41308, by the NSF under CAREER Award No. PHY-0544278, 
by the University of Kansas General Research Fund Program, and by the 
Emmy Noether Program of the Deutsche Forschungsgemeinschaft.
Computations were performed on facilities supported by the NSF under
Grants No. EIA-032078 (GLOW), PHY-0516857 (CMS Research Program
subcontract from UCLA), and PHY-0533280 (DISUN), and by the Wisconsin
Alumni Research Foundation.

\begin{appendix}

\clearpage
\section{More Details on the Optimization of NO$\nu$A*}
\label{app:nova}

\begin{figure}[t!]
\begin{center}
\includegraphics[height=15cm]{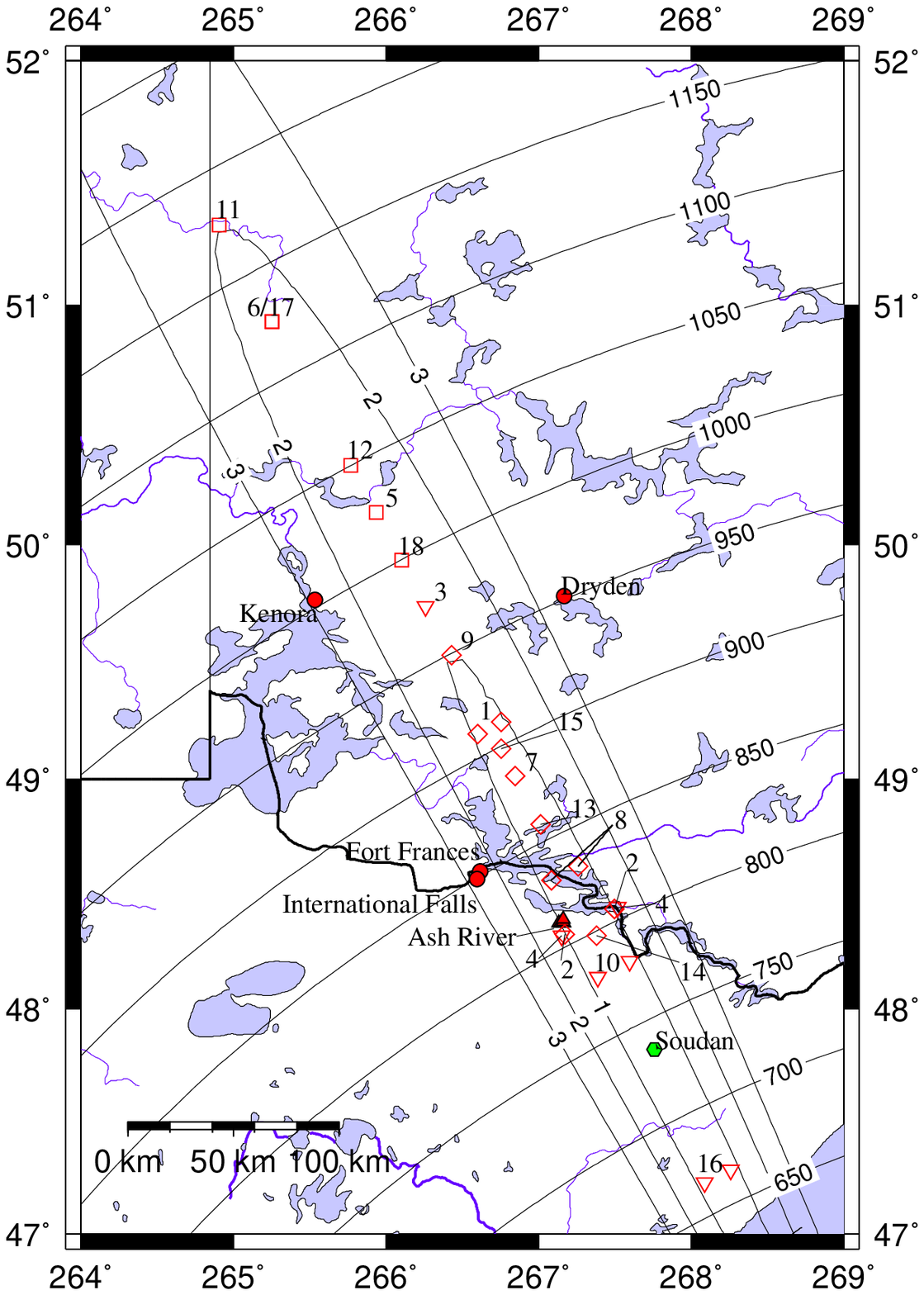}
\end{center}
\mycaption{\label{fig:novageo} Location of the points with optimal
  sensitivity reach in $\stheta$. Shown are all the optima for
  $\deltacp=\pm90^\circ$ and $\Delta
  m^2_{31}=(2.0,2.5,3.0)\cdot10^{-3}\,\mathrm{eV}^2$. Red diamonds
  mark the points for the discovery of $\stheta$, inverted red
  triangles mark the points for CP violation and red squares mark the
  points for discovery of the normal mass hierarchy.  The numbers refer to
  the sites in Tables~\ref{tab:opt20} -- \ref{tab:opt30} where the
  actual achievable sensitivities are listed. The upright red filled
  triangles denote the six possible NO$\nu$A sites given
  in~\cite{Ayres:2004js}. The contours denote places with the same
  off-axis angle (ellipses) in degrees and with same baseline
   from Fermilab (circles) in km.  The thick black curve is the U.S./Canadian
  border. The map is a Mercator projection. }
\end{figure}

In Tables~\ref{tab:opt20}--\ref{tab:opt30} we list all sites which have optimal
sensitivity in at least one of the six cases $\deltacp=\pm90^\circ$
and $\Delta m^2_{31}=(2.0,2.5,3.0)\cdot10^{-3}\,\mathrm{eV}^2$ for at
least one of our performance indicators, and we show the actual
sensitivity reaches.  The geographical location of those sites is
shown in \figu{novageo}.  What this figure demonstrates is that the
Ash River site is the best possible choice within the U.S. and is not
far from the optimum configuration. However, note that many of
the optimal locations are on Canadian soil, especially for the
determination of the mass hierarchy.

\vskip0.25in

\begin{table}[h!]
\begin{center}
\begin{tabular}{||r|rr|rr|rr|rr||}
\hline
\hline
&\multicolumn{2}{|c|}{Site}&\multicolumn{2}{c}{$\stheta$}&\multicolumn{2}{c}{CPV}&\multicolumn{2}{c||}{$\text{sgn}\Delta
  m^2$}\\
&OA $[\circ]$&L $[\mathrm{km}]$&$+90^\circ$&$-90^\circ$&$+90^\circ$&$-90^\circ$&$+90^\circ$&$-90^\circ$\\
\hline

$\stheta^\mathrm{opt}\,[10^{-3}]$ & \multicolumn{2} {|c|} {}  & 0.91 &
2.47 & 10.95 & 2.23 & 41.36 & 55.4   \\
\hline
 1 & 0.91 & 910 & 1.00 & 1.00 & 1.10 & 1.03 & 1.39 & \text{n.s.} \\
 2 & 1.02 & 806 & 1.04 & 1.00 & 1.37 & 1.05 & 1.95 & \text{n.s.} \\
 3 & 1.03 & 975 & 1.11 & 1.19 & 1.00 & 1.18 & 1.14 & \text{n.s.} \\
 4 & 1.14 & 806 & 1.04 & 1.04 & 1.35 & 1.00 & 1.89 & \text{n.s.} \\
 5 & 1.25 & 1025 & 1.77 & 2.50 & 1.08 & 4.97 & 1.00 & 1.75 \\
 6 & 1.7 & 1126 & 15.90 & 28.70 & 2.47 & 21.80 & 1.75 & 1.00 \\
\hline
 7 & 0.62 & 884 & 1.09 & 1.06 & 1.23 & 1.19 & 1.56 & \text{n.s.} \\
 8 & 0.68 & 832 & 1.11 & 1.06 & 1.36 & 1.18 & 1.85 & \text{n.s.} \\
 9 & 0.91 & 949 & 1.00 & 1.03 & 1.03 & 1.04 & 1.25 & \text{n.s.} \\
 10 & 0.68 & 780 & 1.19 & 1.13 & 1.58 & 1.24 & 2.29 & \text{n.s.} \\
 11 & 1.93 & 1176 & 27.00 & 15.00 & 5.89 & 21.80 & 1.25 & 1.39 \\
 12 & 1.36 & 1050 & 2.63 & 4.67 & 1.16 & 8.61 & 1.01 & 1.47 \\
\hline
 13 & 0.57 & 858 & 1.14 & 1.13 & 1.33 & 1.25 & 1.70 & \text{n.s.} \\
 14 & 0.23 & 798 & 1.59 & 1.52 & 1.70 & 1.51 & 2.22 & \text{n.s.} \\
 15 & 0.68 & 899 & 1.05 & 1.00 & 1.17 & 1.13 & 1.48 & \text{n.s.} \\
 16 & 0.68 & 666 & 1.42 & 1.47 & 2.29 & 1.46 & \text{n.s.} & \text{n.s.} \\
 17 & 1.7 & 1126 & 15.90 & 28.70 & 2.47 & 21.80 & 1.75 & 1.00 \\
 18 & 1.14 & 1000 & 1.28 & 1.54 & 1.01 & 1.56 & 1.03 & \text{n.s.} \\
\hline
 AR & 0.89 & 810 & 1.07 & 1.03 & 1.41 & 1.13 & 1.98 & \text{n.s.}\\

\hline
\hline

\end{tabular}
\mycaption{\label{tab:opt20} Relative sensitivity reach at $3\sigma$
  for $\deltacp=\pm 90^\circ$ and $\Delta m^2_{31}=2.0\cdot10^{-3}\,\mathrm{eV}^2$ for various
  sites defined by off-axis angle (OA) and baseline ($L$). The 18 sites in this table have been
  identified to be optimal for one specific measurement and one choice
  of $\deltacp=\pm90^\circ$ and $\Delta
  m^2_{31}=(2.0,2.5,3.0)\cdot10^{-3}\,\mathrm{eV}^2$. The third row
  contains the absolute value of $\stheta$ which was achieved at the best site
  for that case. `n.s.' denotes no sensitivity for $\stheta<0.1$. The absolute sensitivity in
  $\stheta$ can be obtained by multiplying each column with the value
  of $\stheta^\mathrm{opt}$ in the that column. The last row labeled AR
  denotes the Ash River sites.}
\end{center}
\end{table}

\begin{table}[p!]
\begin{center}
\begin{tabular}{||r|rr|rr|rr|rr||}
\hline
\hline

&\multicolumn{2}{|c|}{Site}&\multicolumn{2}{c}{$\stheta$}&\multicolumn{2}{c}{CPV}&\multicolumn{2}{c||}{$\text{sgn}\Delta
  m^2$}\\
&OA $[\circ]$&L $[\mathrm{km}]$&$+90^\circ$&$-90^\circ$&$+90^\circ$&$-90^\circ$&$+90^\circ$&$-90^\circ$\\
\hline
$\stheta^\mathrm{opt}\,[10^{-3}]$ & \multicolumn{2}{|c|}{}  & 0.81 & 2.13 & 8.16 & 2.37 & 12.19 & 34.01   \\
\hline
 1 & 0.91 & 910 & 1.13 & 1.30 & 1.05 & 1.46 & 2.45 & 1.28 \\
 2 & 1.02 & 806 & 1.13 & 1.22 & 1.26 & 1.26 & 3.38 & \text{n.s.} \\
 3 & 1.03 & 975 & 1.60 & 2.23 & 1.04 & 3.29 & 2.28 & 1.38 \\
 4 & 1.14 & 806 & 1.27 & 1.49 & 1.31 & 1.69 & 3.26 & \text{n.s.} \\
 5 & 1.25 & 1025 & 4.18 & 9.28 & 1.30 & 8.11 & 2.59 & 1.11 \\
 6 & 1.7 & 1126 & 16.90 & 9.88 & 7.23 & 9.26 & 1.96 & 1.82 \\
\hline
 7 & 0.62 & 884 & 1.00 & 1.03 & 1.09 & 1.06 & 2.85 & 1.36 \\
 8 & 0.68 & 832 & 1.02 & 1.00 & 1.19 & 1.02 & 3.24 & 2.94 \\
 9 & 0.91 & 949 & 1.19 & 1.45 & 1.00 & 1.77 & 2.28 & 1.10 \\
 10 & 0.68 & 780 & 1.04 & 1.01 & 1.34 & 1.00 & 3.81 & \text{n.s.} \\
 11 & 1.93 & 1176 & 10.20 & 6.32 & 9.69 & 4.92 & 1.00 & 2.75 \\
 12 & 1.36 & 1050 & 7.95 & 16.60 & 1.70 & 11.10 & 3.48 & 1.00 \\
\hline
 13 & 0.57 & 858 & 1.01 & 1.04 & 1.16 & 1.05 & 3.07 & 1.53 \\
 14 & 0.23 & 798 & 1.37 & 1.27 & 1.44 & 1.15 & 3.73 & \text{n.s.} \\
 15 & 0.68 & 899 & 1.00 & 1.03 & 1.05 & 1.08 & 2.74 & 1.30 \\
 16 & 0.68 & 666 & 1.16 & 1.13 & 1.81 & 1.04 & 5.86 & \text{n.s.} \\
 17 & 1.7 & 1126 & 16.90 & 9.88 & 7.23 & 9.26 & 1.96 & 1.82 \\
 18 & 1.14 & 1000 & 2.32 & 4.08 & 1.12& 5.39 & 2.33 & 1.29 \\
\hline
 AR & 0.89 & 810 & 1.03 & 1.06 & 1.24 & 1.07 & 3.46 & 2.91\\
\hline
\hline

\end{tabular}
\mycaption{\label{tab:opt25} 
Similar to \Tab~\ref{tab:opt20} but for $\Delta m^2_{31}=2.5\cdot10^{-3}\,\mathrm{eV}^2$.
}
\end{center}
\end{table}

\begin{table}[p!]
\begin{center}

\begin{tabular}{||r|rr|rr|rr|rr||}
\hline
\hline

&\multicolumn{2}{|c|}{Site}&\multicolumn{2}{c}{$\stheta$}&\multicolumn{2}{c}{CPV}&\multicolumn{2}{c||}{$\text{sgn}\Delta
  m^2$}\\
&OA $[\circ]$&L $[\mathrm{km}]$&$+90^\circ$&$-90^\circ$&$+90^\circ$&$-90^\circ$&$+90^\circ$&$-90^\circ$\\
\hline
$\stheta^\mathrm{opt}\,[10^{-3}]$& \multicolumn{2}{|c|}{} 
   & 0.75 & 1.85 & 6.53 & 2.24 & 8.03 & 21.98   \\
\hline
 1 & 0.91 & 910 & 1.57 & 2.32 & 1.08 & 2.86 & 2.68 & 1.28 \\
 2 & 1.02 & 806 & 1.55 & 2.10 & 1.30 & 2.53 & 3.48 & 2.03 \\
 3 & 1.03 & 975 & 2.92 & 5.40 & 1.17 & 5.13 & 2.59 & 1.06 \\
 4 & 1.14 & 806 & 2.03 & 3.25 & 1.42 & 3.77 & 3.62 & 2.16 \\
 5 & 1.25 & 1025 & 9.95 & 11.30 & 2.39 & 9.63 & 3.14 & 1.17 \\
 6 & 1.7 & 1126 & 7.13 & 4.79 & 9.23 & 3.47 & 1.00 & 3.29 \\
\hline
 7 & 0.62 & 884 & 1.04 & 1.23 & 1.02 & 1.30 & 2.63 & 1.01 \\
 8 & 0.68 & 832 & 1.05 & 1.19 & 1.11 & 1.23 & 3.06 & 1.25 \\
 9 & 0.91 & 949 & 1.75 & 2.77 & 1.05 & 3.33 & 2.51 & 1.12 \\
 10 & 0.68 & 780 & 1.03 & 1.12 & 1.23 & 1.11 & 3.59 & 1.53 \\
 11 & 1.93 & 1176 & 10.50 & 13.30 & 9.69 & 14.70 & 6.12 & 4.28 \\
 12 & 1.36 & 1050 & 14.10 & 9.04 & 4.48 & 9.80 & 2.87 & 1.68 \\
\hline
 13 & 0.57 & 858 & 1.00 & 1.16 & 1.07 & 1.18 & 2.87 & 1.13 \\
 14 & 0.23 & 798 & 1.25 & 1.00 & 1.31 & 1.10 & 3.61 & 1.48 \\
 15 & 0.68 & 899 & 1.10 & 1.32 & 1.00 & 1.47 & 2.46 & 0.94 \\
 16 & 0.68 & 666 & 1.06 & 0.91 & 1.58 & 1.00 & 5.04 & \text{n.s.} \\
 17 & 1.7 & 1126 & 7.13 & 4.79 & 9.23 & 3.47 & 1.00 & 3.29 \\
 18 & 1.14 & 1000 & 5.16 & 9.62 & 1.44 & 7.25 & 2.95 & 1.00 \\
\hline
 AR & 0.89 & 810 & 1.18 & 1.43 & 1.20 & 1.57 & 3.21 & 1.48\\
\hline
\hline

\end{tabular}
\mycaption{\label{tab:opt30} 
Similar to \Tab~\ref{tab:opt20} but for $\Delta m^2_{31}=3.0\cdot10^{-3}\,\mathrm{eV}^2$.
}
\end{center}
\end{table}

\clearpage

\end{appendix}

\bibliographystyle{./apsrev}
\bibliography{Superbeams}

\end{document}